# OccuEMBED: Occupancy Extraction Merged with Building Energy Disaggregation for Occupant-Responsive Operation at Scale

Zhang **Yufei**[1], **Andrew** Sonta[1*]

[1]ETHOS Lab, School of Architecture, Civil and Environmental Engineering, École polytechnique fédérale de Lausanne (EPFL), Pass. du Cardinal 13b (Halle Bleue), 1700 Fribourg, Switzerland

*Corresponding author; e-mail: andrew.sonta@epfl.ch

**Abstract**. Buildings account for a significant share of global energy consumption and emissions, making it critical to operate them efficiently. As electricity grids become more volatile with increasing renewable penetration, buildings must provide flexibility to support grid stability. Building automation plays a key role in enhancing efficiency and flexibility via centralized operations, but its implementation must prioritize occupant-centric strategies to balance energy and comfort targets. However, incorporating occupant information into large-scale, centralized building operations remains challenging due to data limitations. We investigate the potential of using whole-building smart meter data to infer both occupancy and system operations. Integrating these insights into data-driven building energy analysis may enable more occupant-centric energy-saving and flexibility at scale. Specifically, we propose OccuEMBED, a unified framework for simultaneous occupancy inference and system-level load analysis. It combines two key components: a probabilistic occupancy profile generator, and a controllable and interpretable load disaggregator supported by Kolmogorov-Arnold Networks (KAN). This design embeds prior knowledge of occupancy patterns and load-occupancy-weather relationships into deep learning models. We conducted comprehensive performance evaluations to demonstrate its effectiveness across synthetic and real-world datasets compared to various occupancy inference baselines. OccuEMBED always achieved average F1 scores above 0.8 in discrete occupancy inference and RMSE within 0.1-0.2 for continuous occupancy ratios. We further demonstrate how OccuEMBED integrates with building load monitoring platforms to display occupancy profiles, analyze system-level operations, and inform occupant-responsive strategies. Our model lays a robust foundation in scaling intelligent and occupant-centric building management systems to meet the challenges of an evolving energy system.

**Keywords**: Occupant-centric building operation, Occupancy inference, Data-driven energy model, Unsupervised learning, Model interpretability, Kolmogorov-Arnold Network (KAN), Energy signature.

## 1. INTRODUCTION

### 1.1. Background

The operation of buildings accounts for over 30% of global energy consumption and for over 25% of energy-related emissions. Reducing energy consumption in building sector stands out as a top priority for the sustainable energy transition [1]. Meanwhile, the promotion of intermittent renewable energy in recent years causes increasing stress on the electricity grid. In this regard, increasing attention has been paid not just to annual cumulative energy usage, but also short-term building load profiles. This is because the flexibility of building demand profiles can help stabilize the grid by adjusting energy usage in response to supply fluctuations. The global demand response capacity is expected to increase by ten times up to 500 GW by 2050 to achieve net-zero targets, which will be mainly contributed by the building sector [8].

Meanwhile, considering that people spend nearly 90% of their time indoors, the comfort of building occupants is always a primary concern [2]. Indoor comfort encompasses various aspects, including temperature, air quality, and lighting, all of which significantly impact occupant health, productivity, and well-being. However, the heterogeneity of occupants' comfort conditions and behavioral patterns can lead to large variations in final energy use [3]. Balancing building energy targets with the need for occupant comfort is a complex yet essential challenge.

To achieve this balance, large commercial buildings are often equipped with a building automation system (BAS) [4]. These systems can adjust heating, ventilation, air conditioning (HVAC), and lighting systems aiming to optimize energy use while maintaining a comfortable indoor environment. Ideally, the system needs to respond both to varying outdoor conditions and to diverse and dynamic occupant needs. Nevertheless, few buildings currently have implemented dynamic and occupant-centric building operations. The major barrier is the complexity associated with monitoring occupant activity, which can be costly and raise privacy concerns [5]. As a consequence, most existing BASs are operated based on rigid rules and fixed schedules. For example, researchers have observed that energy use in many commercial buildings remained almost unchanged during the pandemic, although occupancy dropped drastically [6].

With increasing attention paid to demand flexibility, promoting occupant-centric building operations can introduce additional challenges. For example, the utility may either directly override BAS control according to their demand side requirements, or they may design dynamic pricing and incentive schemes to ensure the grid-supporting operation is the most economically attractive for building managers [7]. In both cases, the operation of the BAS is less likely to respond to unclear or unavailable occupants' needs under the dominating demand-side requirements. As a consequence, occupants' comfort and satisfaction are likely affected, and the demand flexibility potential may be jeopardized by occupants' lack of acceptance [6]. Such dilemmas are prevalent in office buildings, as office occupants are considered "passive participants" in building management given their limited control over building systems.

In this context, to effectively overcome the barriers to scaling occupant-centric operations with matched progress of demand-flexible operation, we need an efficient tool that is capable of uncovering real-time occupancy conditions, system-level energy consumption, and operational modes across diverse buildings, but that only relies on the most widely available energy metering infrastructure. Research has long established that building energy use reflects occupancy patterns and occupant counts. However, the observed building energy metering is an outcome of complex interactions between occupants, environmental factors, and building system conditions, which can even mix several patterns of multiple systems. Without prior knowledge from existing engineering practices and insights, it is unlikely to extract occupancy from the energy metering in an unsupervised manner, nor to identify system properties and operation modes. Nevertheless, existing approaches still fail to effectively integrate this knowledge into their frameworks.

### 1.2. Main Contribution and Paper Organization

Targeting these gaps, we propose OccuEMBED (Occupancy Extraction Merged with Building Energy Disaggregation). OccuEMBED makes the following contributions:
- A unified data-driven energy analysis framework for simultaneous occupancy inference along with system-level load and operation mode analysis, as depicted in Figure 1.
- Two key innovative components: a probabilistic occupancy profile generator incorporating prior knowledge on occupancy patterns (Section 3.3), and a controllable and interpretable load disaggregator, which embeds prior load-occupancy-weather relationships into deep learning-based models, including the recent advance in interpretable deep symbolic regression—Kolmogorov-Arnold Networks (KAN) (Section 3.4).

As we demonstrate with real-world datasets, OccuEMBED can be deployed as a scalable and adaptive tool offering actionable insights for occupant-centric operation for large building portfolios (Section 4.4).

In the remainder of this paper, we first review existing occupant monitoring and building energy data analysis methods, highlighting the feasibility and current limitation of inferring occupancy and operation patterns solely from building energy metering. Then, we detail the theoretical formulation and implementation details of OccuEMBED, followed by comprehensive performance evaluations of it over synthetic and real-world building datasets against various occupancy inference baselines. We further showcase the deployment of OccuEMBED and discuss its future development to promote occupant-centric operation at scale.

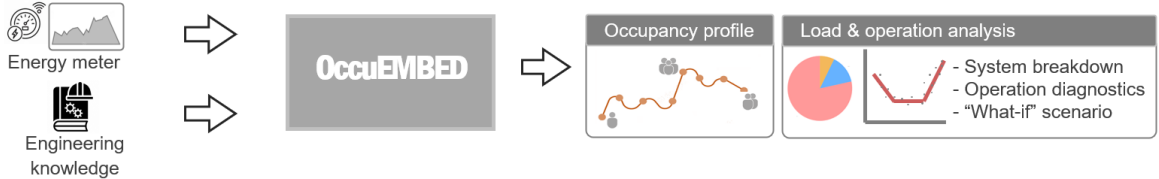

Figure 1 Overview of OccuEMBED.

## 2. LITERATURE REVIEW

### 2.1. *Occupancy Monitoring and Occupant Counting*

Occupant-centric operation requires accurate and efficient methods to obtain occupancy status information. However, there is always a trade-off between better granularity of the occupancy status information and the associated installation and maintenance cost for the sensors, alongside potential privacy invasion [5]. For building-level energy activities, such as energy reduction and demand response, it is sufficient to have a binary occupancy status (occupied/unoccupied) and occupant counting, as it already fits building-level system operation [8].

Nevertheless, this basic information still requires dedicated occupancy sensing systems, such as cameras and radio frequency identification (RFID) tags, which can help to consider the occupant experience in building management but may raise costs and privacy issues [9]. An alternative method leverages inverse modeling, which uses indoor environmental sensing signals from the BAS, such as $CO_2$ concentration and humidity, to estimate occupant numbers [10], [11] [12]. While this method can offer high accuracy, its dependency on sensor availability and exhaustive data collection—including building thermal characteristics, system conditions, and ground-truth occupant counts over a period of time—significantly hinders its scalability and urban-scale applications. Wi-Fi infrastructure, which is widely equipped in commercial buildings, provides a promising opportunity for virtual sensing of occupant count [8], [13], [14]. Nevertheless, the accessibility of Wi-Fi data remains limited. Figure 2 summarizes the accessibilities of relevant data sources for occupant-centric operation under different granularities. The effort to access granular occupant sensing data and BAS indoor sensing data is only worth it for zone and system level operations. Meanwhile, most of the existing studies of occupancy prediction focus on supervised learning that requires ground-truth occupant counts from these data sources, which are difficult to obtain for large-scale whole-building operation, such as demand flexibility services [15], [16].

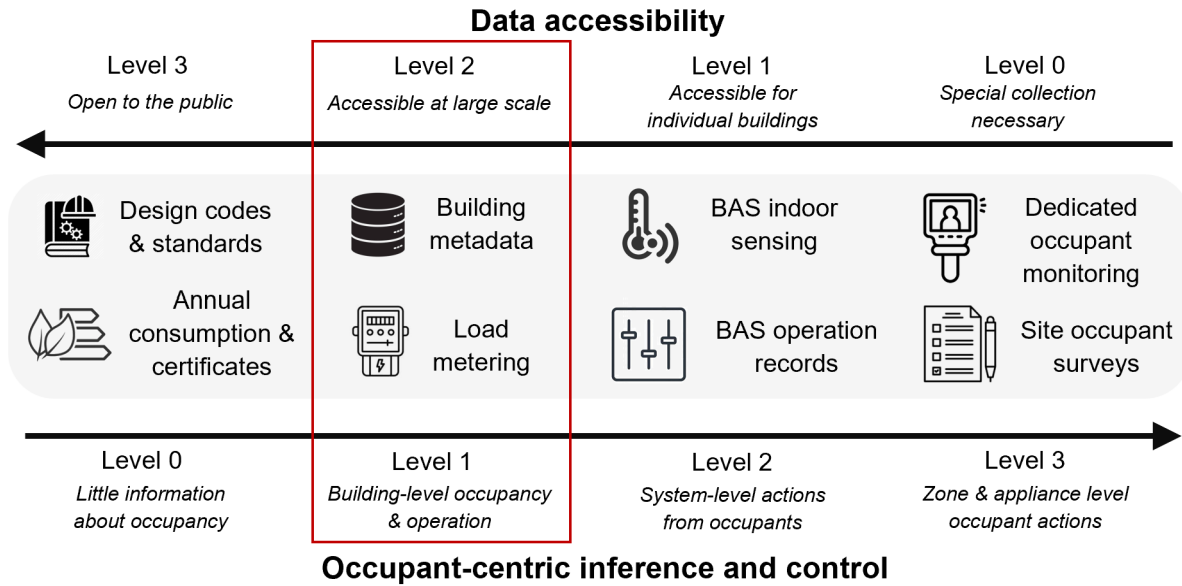

Figure 2 Overview of data sources for occupant-centric inference and operation: accessibility levels vs. operation granularity levels.

*2.2. Occupancy Inference from Energy Metering*

At the same time, electricity metering data, as the most fundamental infrastructure in energy services [17], [18], contain insights into occupancy dynamics that are sufficient for building-level operation, as shown in Figure 2. Patterns of electricity use have varied profiles on working days, weekends, holidays, and between working and non-working hours. These patterns indicate fundamental differences in the underlying human activity patterns [19], [20]. Prior studies have explored unsupervised extraction of occupancy levels and activity types from electricity metering data via latent variable models. For example, Sonta et al. leveraged Gaussian Mixture Model (GMM) to identify different states for individual occupants in two open-space offices from their desk-level plug loads. The activity states include: absent, low activity, and intensive activity [21]. Kleiminger et al. adopted a Hidden Markov Model (HMM) to identify underlying binary occupancy states that cause observed total electricity metering in five households in Switzerland [22]. These prior studies were established on the assumption that the magnitude and variance of the energy consumption are associated with the presence/absence of the occupants, which is usually held in small residential buildings and in fine-grain metering of occupant-driven loads in office buildings such as individual plug loads. However, for most large office buildings, only lumped-metered energy data of an entire building is commonly available, occasionally with submetering by types of systems [6]. Due to the operation modes of the BAS and occupants' energy use behaviors, the energy consumption levels may not be explicitly associated with the level of occupancy when observing the whole-building metering.

To deal with building-level energy data, one related research area is Non-Intrusive Load Monitoring (NILM), which utilizes data-driven models to disaggregate load metering data by appliances and can therefore offer insights into occupants' behavioral patterns. However, NILM methods typically rely on high-frequency load metering and require labeled datasets of appliance operation signatures [23]. When only low-frequency interval metering of electricity use is available, an early stage study suggested that a simple linear transformation of the occupant-driven loads (i.e., lighting and plug loads), can produce a good proxy of the occupancy ratio [24]. However, it is not clear if the inferred occupancy profiles were validated under real building data. Wang et al. highlighted the importance of basic submetering that separates central HVAC plants from other end-use energy systems for office occupants using real building data [25]. Specifically, they found that lighting energy use is associated with the schedule of binary occupancy, and that plug loads could be a very informative feature for the prediction of occupant counts by supervised training. Nevertheless, this study did not discuss how to infer occupancy when ground-truth occupant counting is not available. On the other hand, O'Brien reviewed the occupant-centric operation modes on real-world office buildings and stressed the large proportion of lighting and plug loads during low-occupancy and unoccupied periods, which potentially hinders inferring actual occupancy [26].

In practice, for the inference of occupancy, the form and availability of energy sub-metering might still be problematic. Tenant sub-metering may be required by municipalities [27], [28], since it helps ensure owners or tenants only pay for the amount of electricity they use. Each tenant or unit owner can be individually billed for the actual amount of electricity used, which can provide an incentive for occupants to reduce their electricity usage. Sub-metering for large and important equipment, such as chillers and boilers, might also be available to monitor their operation [29]. However, perfect sub-metering by each energy system might not be easily accessible [18]. Some studies tried to address more limited data availability when there is no submetering at all. Many studies focus on normalizing methods, such as Energy Signatures (ES) that remove weather-induced variation on HVAC energy use, to isolate the impacts of BAS operation and occupants' behavior [30], [31], [32], [33]. Capozzoli et al. developed a data-analysis pipeline that segments daily electricity consumption profiles in office buildings into several intervals according to variance levels. These intervals implicitly represent unoccupied periods, ramping up and down in the morning and evening, as well as peak occupancy periods during the day [34]. Nevertheless, this coarse information may fail to reflect the subtle inter-day variations of patterns, thus restraining sufficiently occupant-responsive operations. Rouchier proposed an innovative workflow that combines ES with HMM for modeling heterogeneous transitions among occupancy levels over periods of a day to account for impacts of occupancy on load patterns. This Bayesian inference-based framework enables a probabilistic description of energy use and occupancy, allowing the incorporation of prior knowledge to improve model training. However, the large number of parameters to be identified through Bayesian inference requires users to have strong expertise in both statistical modeling and building energy analysis, potentially limiting its accessibility to practitioners. Furthermore, there is no direct validation of the accuracy of its occupancy inference [35]. We developed preliminary work to infer occupancy as a hidden state and disaggregate system-level loads from smart meter data using Variational Autoencoders (VAE), though this initial work lacked the incorporation of prior knowledge to guarantee realistic occupancy profiles and shortcomings with handling HVAC loads [36]. Opportunities to address these shortcomings of existing work include Variational Quantized-Variational Autoencoder (VQ-VAE) [37], where the use of pre-generated samples to define the latent space can enable the implicit incorporation of plausible occupancy patterns, and KAN [38], which advances interpretable deep symbolic regression, allowing to integrate ES to handle weather-driven HVAC loads.

Overall, the most scalable data source for supporting occupant-centric operation is whole-building energy metering. Although prior studies identified that information on occupancy and occupant counts is reflected by whole-building energy use, there remains a need to design a system to recover occupancy from this energy data credibly and robustly. Doing so requires a systematic strategy that incorporates explicit definitions of the relationships between occupancy, environmental factors, and system-level loads—enabling the disentanglement of their complex interactions. This is our primary aim in this paper.

## 3. METHODOLOGY

### 3.1. Model Overview

In this work, we propose a modeling framework to infer building occupancy levels and system-level operation patterns from observed energy metering data. This task addresses two key challenges:
1. **Hidden variable inference:** estimating unobserved (hidden) variables $Z$ (occupancy levels in our case) from observed data $P$ (power load monitoring in our case).
2. **Incorporation of prior knowledge:** integrating domain knowledge, such as typical daily occupancy patterns and the influence of exogenous variables $X$ (weather conditions in our case) into the inference process. Effective incorporation of prior knowledge is critical, as the load-occupancy-weather relationships vary across different building systems, each characterized by distinct model structures and parameters $\theta$. By capturing these relationships, we can also disaggregate whole-building loads by system type and gain fundamental insights into system-level operation modes.

From a statistical learning perspective, this task falls under the classical hidden state inference problem, where the goal is to estimate the posterior distribution of the hidden state, $p(Z|P, X; \theta)$, while accurately identifying the associated model parameters $\theta$ via Maximum Likelihood Estimation (MLE). Appendix A provides a detailed formulation of the problem within this context. We also elaborate the inspiration from existing hidden state inference methods—Expectation-Maximization (EM) algorithm and VAE—to address the hidden state problem and highlight the special challenges in our case.

Our framework is summarized in Figure 3. It incorporates two key designs:

1. **A probabilistic occupancy profile generator (Section 3.2):** the generation of stochastic probabilistic occupancy profiles is based on reference schedules derived from local design guidelines and surveys. By introducing controlled disturbances to these reference schedules, the generator ensures a broad yet plausible solution space for candidate posteriori occupancy profile distributions, $p(Z|P, X; \theta)$. This ensures that the inference process remains consistent with known occupancy behaviors while being adaptable to building-specific conditions.

2. **A controllable and interpretable load disaggregator (Section 3.3):** the load disaggregator incorporates explicit mathematical links between energy metering, occupancy levels, and weather conditions. These mathematical links are structured based on prior knowledge of how various energy systems—such as lighting, plug loads, and HVAC systems—respond to occupancy and external factors like weather. By parametrizing these distinct links via different sets of parameters $\theta$, the model disaggregates whole-building energy metering into system-level loads in an interpretable manner. To enhance this process, we integrate the advances of deep symbolic regression—KAN, along with specialized designs for heteroscedastic regression to account for subtle variability in load patterns.

The entire workflow is implemented in a deep learning framework to enable hardware acceleration and streamlined training pipeline. Since neither the occupancy profiles nor the load disaggregator parameters are known a priori, we employ an alternating training process that iteratively refines both components, ensuring convergence toward an optimal solution. The model is trained using a two-step process leveraging EM applied to our scenario (gray boxes in Figure 3; described in Section 3.4).

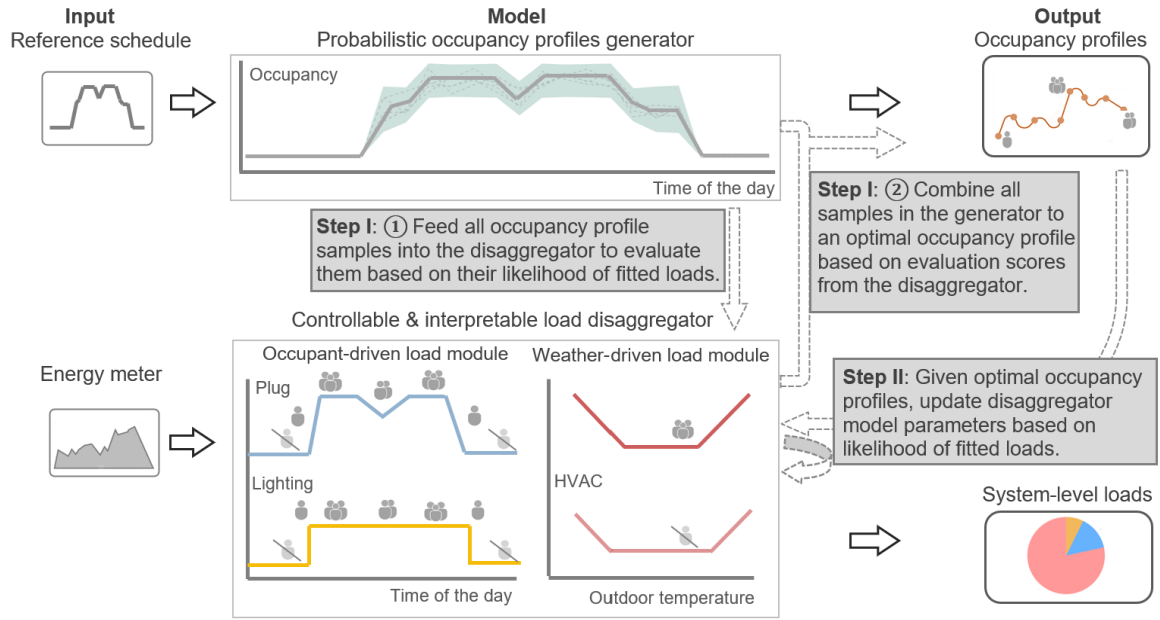

Figure 3 OccuEMBED architecture and its alternating training steps. The solid arrows depict the input / output pipeline in deployment stage. The dashed arrows model internal data flow during the training process.

### 3.2. *Probabilistic Occupancy Profile Generator*

*3.2.1. Stochastic Occupancy Profile Samples*

Office building occupancy patterns exhibit complex intra-day variations, reflecting the diurnal rhythm of social activities. Directly inferring the probability distributions of occupancy levels at different times of the day can be problematic, as it may lead to deviations from typical office occupancy patterns. To address this, we take inspiration from VQ-VAE [37] and generate a finite set of candidate samples to represent the solution space of time-indexed posterior occupancy distributions $p(Z_t|P_t, X_t)$. This approach avoids the difficulty to explicitly calculate the posterior distribution and ensures the inferred time-indexed posterior distributions follow realistic occupancy patterns aligned with reference schedules.

To ensure validity and practical utility, these samples are designed to satisfy several essential properties:
1. Each sample must represent a valid probability distribution, meaning that the probabilities assigned to all levels sum to 1.
2. Each sample distribution's expectation must fall within the range [0,1], capturing all plausible occupancy ratios, from no occupancy (0) to full occupancy (1).
3. The distribution should have a dominant mode that highlights the most likely occupancy level at a given time. This ensures the sample reflects uncertainty while also indicating the most probable occupancy states.

A practical choice that satisfies these conditions is the categorical distribution, which discretizes the continuous range of occupancy ratios [0,1] into distinct bins. Each bin represents a specific occupancy levels, allowing the distribution to be valid, flexible, with distinct modes. Additionally, to ensure full coverage of the [0,1] range, two special boundary levels are always included: 0—representing non-occupancy during inactive periods; and 1—representing full occupancy. For example, a categorical distribution with four levels may use level centroids at [0, 1/3, 2/3, 1].

Forming discrete categorical distribution also facilitates the generation of stochastic occupancy profile samples based on the given reference schedules. As illustrated in Figure 4a, we compare the reference fractional occupancy ratio at each time step with the specified level centroids, $c_i$, to identify the most probable discrete occupancy level. We then assign logit values to each level and apply a SoftMax transformation to construct categorical distributions as probabilistic occupancy profile samples. Appendix B provides further details on the sampling mechanism for input logit values and the generation of multiple probabilistic occupancy profile samples with discrete occupancy levels. It also includes a discussion of how the dispersion degree of the categorical distributions can be controlled to introduce different levels of uncertainty in the occupancy profile samples. Figure 4b presents an example of how probabilistic occupancy samples at each hourly step form a daily occupancy profile.

However, the observed load output is still continuous. It is impractical to establish the mathematical link between the discrete random variable of occupancy levels and continuous load output. To address this limitation, we introduce a Gaussian Mixture (GM) distribution as a proxy for the continuous fractional occupancy ratio for each generated sample of discrete levels. Specifically, the discrete levels are already defined as bins and boundaries within the range [0,1]. Based on this structure, we fix the mean and variance of the Gaussian component corresponding to each level's centroid. The weight of each component is determined by the probability mass of that level in the categorical distribution. Here, we need to treat discrete bin levels and boundary levels differently:
- For the discrete bin levels, the centroids of the bins serve as the means of the corresponding Gaussian components, while the variance of a uniform distribution over the bin is used as the variance for the Gaussian components.
- For the boundary levels (0 and 1), we set the means of the corresponding Gaussian components with small offsets from the boundaries. Additionally, small variances are also assigned to minimize violations of the valid range [0,1]. While in this case, boundary violations cannot be entirely avoided, they are negligible and acceptable due to the probabilistic nature of occupancy ratio samples.

Figure 4b and c illustrate the transformation process from a categorical distribution to its GM proxy. This transformation enables the load output of each system to be modeled as a continuous random variable derived from occupancy ratios, also following GM distributions. This property will be discussed with more details in Section 3.3.

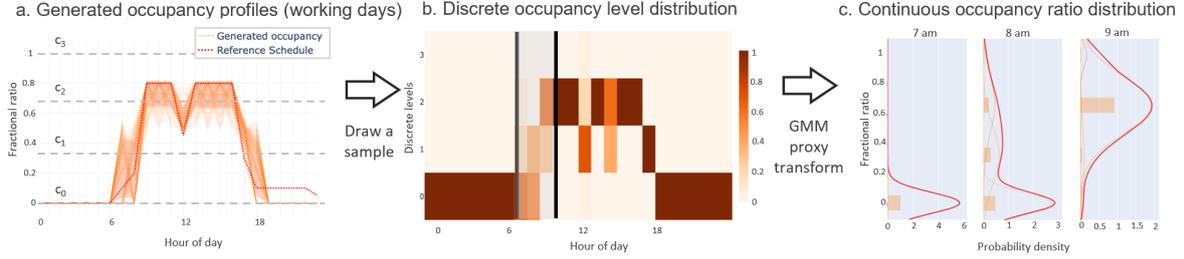

Figure 4 Illustration of the probabilistic occupancy sample generation mechanism. (a). Reference (ASHRAE typical office) and generate daily occupancy ratio profiles (mean value) with pre-defined level centroids. (b). An example of daily discrete level probabilistic occupancy profile. (c). The GM proxy for probabilistic continuous occupancy ratio by hourly steps.

### 3.2.2. Constructing Optimal Occupancy Profiles

The generated occupancy profile samples constitute a plausible solution space of time-indexed candidate posterior distributions $p(Z_t|P_t, X_t)$. This means we that the given load output $P_t$ and the exogenous variable $X_t$, we can evaluate each candidate sample based on the output likelihood and construct the optimal distribution by combining the samples with the highest likelihood, as described below.

The process is implemented in a sequential manner. For each candidate series of posterior distributions $\widehat{Z_{t_1:t_2}}$, we feed it into the load disaggregator and get its output log-likelihood, which measures how well the candidate matches the observed load output sequence $P_{t_1:t_2}$. Its log-likelihood is expressed as the summation of log-likelihood over the time steps $\mathcal{L}(\widehat{Z_{t_1:t_2}} \mid P_{t_1:t_2}) = \sum_{t=t_1}^{t_2} \log p(P_t \mid \widehat{Z_t}, X_t, \theta)$. The exact formulation of the log-likelihood will be detailed in Section 3.4.1.

Next, we calculate a matching score $w^{\text{match}}(\widehat{Z_{t_1:t_2}})$ for each candidate series $\widehat{Z_{t_1:t_2}}$ by applying a softmax transformation to its log-likelihood value, as shown in Equation (1). This scoring mechanism, inspired by cross-entropy-based sampling methods [39], quantifies the likelihood of each candidate posterior series may contribute to the actual underlying posterior distribution. These scores are then used as weights to compute a weighted average over the candidate samples to get the optimal posterior distributions. Here, $N$ represents the number of selected candidate posterior samples. To avoid over-smoothing the optimal posterior distribution by including low-score candidates, we restrict the computation to a small subset of samples with the top-$K$ scores.

$$w^{\text{match}}(\widehat{Z_{t_1:t_2}}) = \frac{exp\left(\mathcal{L}(\widehat{Z_{t_1:t_2}} \mid P_{t_1:t_2})\right)}{\sum_{k=1}^{K} exp\left(\mathcal{L}(\widehat{Z^N_{t_1:t_2}} \mid P_{t_1:t_2})\right)}$$

(1)

### 3.3. Controllable and Interpretable Load Disaggregator

The generated posterior distribution at each step $p(Z_t|P_t, X_t)$, as a GM distribution, are fed into the load disaggregator. The load disaggregator transforms $Z_t$ and the exogenous variables $X_t$ into system-specific loads $P_{sys,t}$. This transformation ensures that both system-level outputs $P_{sys,t}$ and their summed total building load $P_t$ preserve as GM random variables. This property allows us to design loss enables the parameters of the load disaggregator to be fitted using MLE of the observed $P_t$.

We detail the design of the load disaggregator by explaining how its architecture incorporates prior knowledge and how its parameters align with actual physical interpretations. Additionally, we discuss how the GM distribution is preserved throughout the transformations. Building system loads are categorized into two groups—occupant-driven and weather-driven—each modeled with a dedicated module, as depicted in Figure 5.

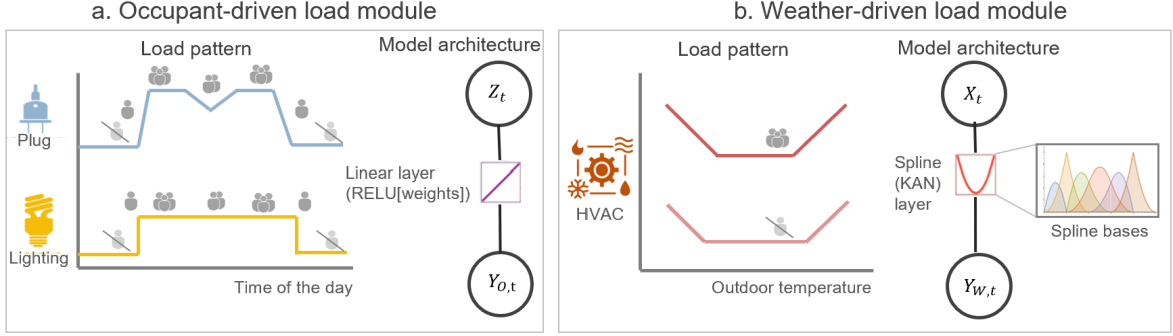

Figure 5 Load disaggregator module overview by illustration of load patterns and diagram of model architectures. (a) Occupant-driven load module. (b) Weather-driven load module.

*3.3.1. Occupant-Driven Load Module*

In office buildings, internal lighting and plug loads are considered occupant-driven because they directly depend on occupancy conditions. As many prior field and simulation studies demonstrate [20], [25], [40] and how our initial work implemented [36], we can link the GM proxy variable of occupancy $Z_t$ and occupant-driven load $Y_{o,t}$ with an affine equation:

$$P_{o,t} = P_{dynamic} \cdot Z_t + P_{base}$$

(2)

Here, $Z_t$ is the only input into the module, while $P_{dynamic}$ and $P_{base}$ are the module parameters. $P_{dynamic}$ represents the portion of the system's installed capacity that varies dynamically with occupancy, such as lighting or plug loads in office areas. Meanwhile, $P_{base}$ accounts for constant energy use from devices that operate independently of occupancy, such as data storage units. To ensure their physical interpretability as system capacities, both $P_{dynamic}$ and $P_{base}$ are constrained to be non-negative. In the deep learning framework, this constraint is implemented by defining a linear layer and applying a Rectified Linear Unit (ReLU) to the weight and bias, ensuring they correspond to $P_{dynamic}$ and $P_{base}$ respectively, as the diagram in Figure 5a depicts. In addition, prior knowledge can be incorporated into the initial values of these parameters to improve convergence and model performance. For instance, if the building's floor area and energy certification category are available in a building metadata database, the average power intensity for that category can be multiplied by the floor area to estimate the initial installed capacity.

We further disaggregate lighting and plug loads. Their key distinction is the responses to nuanced representations of occupancy conditions. As illustrated in Figure 5a, plug loads are more dependent on occupancy. In Equation (2), $Z_t$ directly corresponds to the continuous occupancy ratio, meaning dynamic plug loads are proportional to the occupancy ratio. In contrast, lighting loads are much less dependent on small variations in occupancy. Lighting for a large space may switch on entirely as long as one occupant is present. In this case, $Z_t$ is better represented as a binary occupancy variable, indicating the probability of occupied versus non-occupied states. This approach appears to conflict with the original GM proxy, which represents occupancy ratio as a continuous range. However, we can approximate binary occupancy by overwriting all Gaussian components, except the zero-occupancy component, with the mean and variance of the full-occupancy component. This adjustment preserves the weights of each component, allowing the GM proxy to approximate a binary occupancy distribution. Finally, because the GM distribution is preserved under affine transformations, the lighting and plug loads still follow GM distributions. Additionally, the load output of different systems may have distinct component means and variances. Still, their component weights remain consistent with the probability mass derived from the discrete occupancy-level categorical distribution. As a result, the sum of the output from different systems follows a new GM distribution with the same component weights. Each

Gaussian component in this new GM distribution has a mean and variance equal to the sum of the means and variances of the corresponding components in the original distributions.

*3.3.2. Weather-Driven Load Module*

The other major category of building energy systems consists of loads primarily driven by outdoor weather conditions. The most significant system in this category is the HVAC system. Overall, HVAC system loads result from the combined effects of building hygrothermal dynamics, HVAC system behavior, and occupant activity. Capturing these complex dynamics is difficult with only whole-building energy metering and minimal information about building conditions. Therefore, a practical alternative is to focus on the primary weather factor: outdoor air temperature. Our approach focuses on outdoor air temperature and captures the static heating and cooling supply, which dominates the operating time in automatically controlled office buildings. Despite this simplification, the relationship between HVAC loads and outdoor temperature often exhibits complex piecewise patterns:

1. Building heat loss and gain are typically driven by a constant factor multiplied by the temperature difference, resulting in a linear relationship. However, as outdoor temperatures change, the load output may transition between cooling, heating, and intermediate periods. This can also reveal behaviors such as multi-chiller/boiler activation or insufficient system sizing under extreme conditions. As a result, the slopes of the load-temperature relationship differ across these ranges, as illustrated in Figure 5b.
2. Moreover, building-specific factors, such as envelope properties, system types, and operational modes, introduce varying patterns with distinct slopes and change points, posing challenges for application across a large portfolio of heterogeneous buildings.
3. Finally, occupancy conditions also significantly influence HVAC loads. For example, HVAC systems often follow schedules with indoor temperature setpoints and setbacks that differ between occupied and unoccupied periods. Internal gains from occupant-driven loads further complicate the load-temperature relationship. These occupancy factors potentially create bifurcations in the pattern.

The classical ES method addresses the challenge of modeling the relationship between energy consumption and outdoor air temperature. As discussed in Section 2.2., prior studies have proposed using flexible piecewise linear or spline models to account for building heterogeneity [31]. These approaches enable a single model setup to scale across diverse building cases. Other works also suggested ensuring different parameter sets in occupied/non-occupied periods [35]. While these results are insightful, they were developed primarily for whole-building load analysis rather than system-level disaggregation. In the context of the load disaggregator, we should implement a suitable ES model within a deep learning framework side-by-side with the occupant-driven load module. This pipeline ensures that system-specific loads are automatically disaggregated during training.

To implement the weather-driven load module, we adopt the recently proposed KAN, a specialized neural network designed for interpretable scientific modeling [38]. Unlike standard neural networks, which learn complex patterns by "deeply" stacking multiple linear layers and simple activation functions, KAN focuses on capturing the complex activation of individual inputs using "shallow" layers. This makes it particularly suitable for identifying explicit equations in scientific modeling. KAN also offers high-level interpretability by fitting activations over a set of scientifically meaningful functional bases, particularly flexible B-spline bases. This design makes KAN well-suited for our ES model, as its spline layer can effectively capture diverse ES patterns across multiple buildings, as illustrated in Figure 5b.

To account for the bifurcation of ES patterns between occupied and unoccupied periods, we use two parallel KAN spline layers with distinct parameter sets, as shown in Equation (3). In this formulation, $\beta_i(Z_t)$ represents the gate selection over binary occupancy levels, which determines the activation of the respective KAN spline layer—denoted as $B_i$ for either the zero occupancy or full occupancy levels. Thus, as in the lighting load module, we approximate the binary occupancy levels using the continuous GM distribution. Also similarly, the weather-driven HVAC load $P_{w,t}$ follows a GM distribution, where the component weights remain unchanged, and the component means are shifted as the KAN spline output. Finally, the total building load $P_t$, obtained by summing the system-level outputs $P_{o,t}$ and $P_{w,t}$ also follows a GM distribution.

$$P_{w,t} = \sum_{i=0}^{1} \beta_i(Z_t) B_i(X_t)$$

(3)

### 3.4. Implementation

#### 3.4.1. Loss Function Design

As explained in Section 3.3, the observed load output $P_t$ follows a GM distribution. The likelihood of the observed $P_t$ provides a key reference for constructing the optimal posterior occupancy $p(Z_t|P_t, X_t)$. It also serves as the foundation for applying MLE to estimate the load disaggregator parameters $\theta$.

First, to select candidate surrogate posterior occupancy samples and combine them into the optimal distribution based on matching scores $w^{\text{match}}(\widehat{Z_{t_1:t_2}})$. We calculate the likelihood of the observed $P_t$. This is done using the GM distribution defined by each candidate $\widehat{Z_{t_1:t_2}}$. In this process, we are assuming each candidate sample reflects the true underlying occupancy distribution. As a result, we use the full log-likelihood formulation for the GM proxy:

$$\mathcal{L}_{generator} = -\log \sum_{t=t_1}^{t_2} \sum_{k=1}^{K} \pi_{k,t} \cdot \mathcal{N}(P_t|\mu_{k,t}, \sigma_{k,t}^2; \theta)$$

(4)

Once the optimal surrogate posterior is constructed, MLE is applied to estimate the load disaggregator parameters. When the distribution of $Z_t$ is fixed, the objective simplifies to the expectation of the joint log-likelihood over all the posterior distributions. We then take the negative form as the loss function to be minimized, as shown in Equation (5). The detailed derivation is similar to that in the classical EM algorithm, which is provided in Appendix A:

$$\mathcal{L}_{disaggregator} = -\sum_{k=1}^{K} \pi_{k,t} \cdot \log \mathcal{N}(P_t|\mu_{k,t}(\theta), \sigma_{k,t}(\theta)^2)$$

(6)

This equation also highlights that the load disaggregation parameters determine the means and variances of the Gaussian components, resulting in heteroscedastic behavior. While this setup is more realistic than the common homoscedastic assumption, which overlooks potential higher variability in high-load ranges, it poses challenges for training. The model may tend to arbitrarily inflate variance estimates instead of properly fitting the mean values. To address this issue, we apply the $\beta$-NLL trick [41], which modifies the log-likelihood term by weighting it with the $\beta$-exponentiated variance estimate, as shown in Equation (7). This adjustment penalizes overly large variance estimates, promoting better parameter fitting.

$$\mathcal{L}_{\beta-NLL} = \sum_{k=1}^{K} \pi_{k,t} \cdot (\sigma_{k,t}(\theta)^2)^{\beta} \cdot \log \mathcal{N}(P_t|\mu_{k,t}(\theta), \sigma_{k,t}(\theta)^2)$$

(8)

#### 3.4.2. Training Process

As illustrated in Figure 3, our model follows an alternating two-step training process:
- Step I: Using the current load disaggregator parameters, first, we evaluate all candidate occupancy profile samples by feeding them into the load disaggregator and computing their log-likelihood scores based on the observed load output. These scores are then used to combine the samples, constructing the optimal occupancy profiles for observed load output. The details of this process are provided in Section 3.2.2.

- Step II: With the optimal occupancy profiles obtained from Step I, we update the parameters of the load disaggregator to better fit the observed load output.

This alternating approach iteratively refines both components, which is analogous to the alternating Expectation-step and Maximization-step in the EM algorithms. On the other hand, our model avoids directly computing the posterior occupancy distribution by leveraging the predefined solution space of candidate samples. This aligns with the principle of variational inference in VAE models. Appendix A elaborates on the connections between these established hidden state inference methods and our approach. Drawing inspiration from these established methods, OccuEMBED demonstrates robust and stable performance in experiments despite no theoretical convergence guarantees. We implemented our model in the widely-used deep learning framework PyTorch [42]. Typically, the model converges within 5 to 10 epochs, provided that historical energy metering data spanning over several months to a year is available.

We also found that a total sample size of approximately 2,000 candidates (1,500 for working days and 500 for non-working days) is sufficient for robust performance. We note that the variety of candidate samples, rather than their sheer quantity, is critical to model performance. This aspect is discussed in greater detail in the discussion on model limitations below.

## 4. RESULTS: PERFORMANCE BENCHMARKING

### 4.1. Evaluation Strategy

#### 4.1.1. Baseline Methods

To comprehensively benchmark OccuEMBED's performance, we selected a diverse set of existing unsupervised occupancy inference methods as baselines. Most existing methods rely on clustering applied to observed load data. These methods only infer discrete occupancy levels. We implemented the following baselines:

1. K-means: a widely-used clustering algorithm due to its simplicity and efficiency. It partitions data into $K$ clusters by minimizing within-cluster variance. Its iterative optimization process, which iteratively identifies clustering centroids and updates allocation of the data to each cluster, resonates with the EM algorithm's concept of alternating updates. However, the main drawback of K-means is the implicit assumption that the identified clusters, as inferred occupancy levels, are homogeneous all the time. We implemented it using the Scikit-learn Package [43].
2. GMM: a probabilistic clustering method that models the data as a mixture of Gaussian distributions. Like K-means, it leverages EM for iterative parameter estimation but allows for soft assignments of data points to each Gaussian component. Similarly, GMM also assumes homogeneous distribution of the underlying occupancy levels. We implemented it using the Scikit-learn Package [43].
3. HMM with input variables: a statistical hidden state model that accounts for temporal dependencies by assuming hidden states (i.e., underlying occupancy levels) govern observed outputs, with transitions between states following a Markov process. In this HMM implementation, we also incorporate the most critical exogenous variables, hour of the day and day type, into the state transition probabilities. This property potentially overcomes the drawback of K-means and GMM and allows heterogeneous distributions and transitions of the underlying occupancy levels across different periods of the day. The initial transition matrix reflects prior knowledge of typical low-occupancy, high-occupancy, and transitional periods. Nevertheless, this design also significantly increases the number of parameters to be fitted, and there is still no explicit prior knowledge incorporation as in our model. We implemented it using the Pomegranate package [44].

Additionally, we implemented a continuous occupancy ratio inference using the linear transformation of occupant-driven loads proposed in [24]. Similar to the logic in Equation (2), this transformation is based on the assumption that the dynamic portion of the load $\frac{(P_t - P_{min})}{(P_{max} - P_{min})}$ is proportional to the occupancy ratio $Z_t$. Here, $Z_{max}$ is a tunable hyperparameter reflecting our estimation of the maximum possible occupancy ratio:

$$Z_t = Z_{max} \frac{P_t - P_{min}}{P_{max} - P_{min}}$$

(9)

*4.1.2. Evaluation Processing Pipeline and Metrics*

It is important to note that we evaluate our model under two different scenarios of energy metering aggregation levels:
1. **Separate**: in this scenario, occupant-driven loads (e.g., lighting and plug loads) are metered separately at the office unit level, while weather-driven loads are derived from central HVAC plant metering. This scenario is available in some modern multi-tenant office buildings equipped with advanced submetering systems. This is an easier and more desired scenario for occupancy inference.
2. **Lumped**: in this scenario, both occupant-driven loads and weather-driven HVAC loads are lumped into a single whole-building metering record. This is a more common scenario for utilities considering a large portfolio of buildings.

A significant challenge for all baseline methods arises from the lumped weather-driven loads, which breaks the simple assumption that higher loads indicate higher occupancy. To mitigate this issue, referring to the established works in [31], we apply a weather-trend removal process. This involves fitting a piecewise linear ES model with optimal breakpoint identification to the whole-building load data and removing the weather-driven trend from the raw data, as depicted in Figure 6.

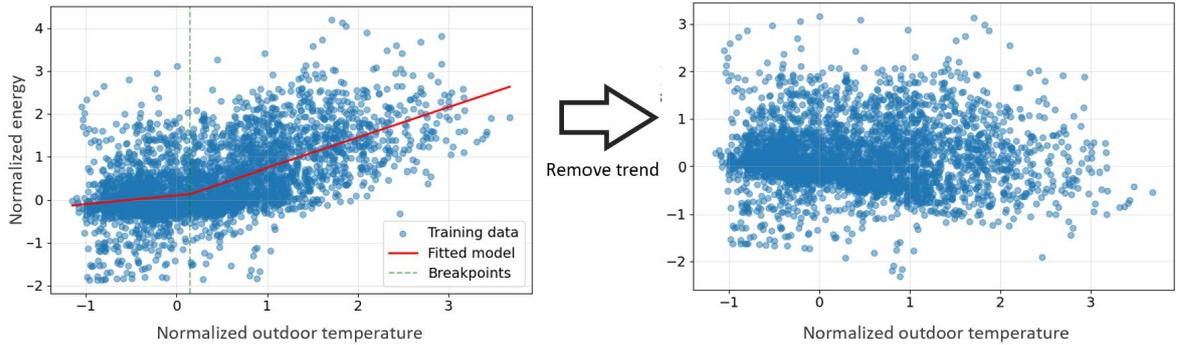

Figure 6 Illustration of weather-trend removal based on breakpoint piecewise linear ES method.

Although HMMs could theoretically model the impact of outdoor temperature in the emission process [35], experiments revealed that introducing piecewise linear functions into HMMs greatly increased parameter complexity and hindered performance. As a result, we apply weather-trend removal to all baseline methods before fitting them to the data. This preprocessing step ensures more competitive baseline results.

For clustering methods, we evaluate performance using the F1 score, which measures precision and recall for each discrete occupancy level. Since ground-truth occupancy is available as a continuous ratio, we apply K-means clustering to partition the ground-truth occupancy into discrete levels. Note that this direct processing of the ground-truth occupancy is just for evaluation. No ground truth was provided to train the unsupervised occupancy inference models. The predicted occupancy ratios from our model are similarly categorized into these levels. We then compute the F1 score for each level, as well as the macro and weighted averages of the F1 scores.

Since clustering methods are unsupervised, their optimal number of clusters may differ from the predefined three levels. We observed that setting this number to 5 or 6 often produces the best fit. To ensure optimal performance of the baselines, we manually map these additional clusters to the three occupancy levels until the highest macro and weighted average F1 scores. While this approach is unrealistic in real-world scenarios—where ground-truth occupancy is unavailable—it helps establish the most competitive baselines for comparison.

For continuous ratio methods—the linear transformation baseline and our OccuEMBED—we directly evaluated performance using Root Mean Squared Error (RMSE). This metric directly compares the predicted and ground-truth occupancy ratios. We calculate RMSE for each of the three abovementioned occupancy levels (low, medium, high) and report the overall RMSE.

Since $Z_{max}$ is an essential hyperparameter in the linear scaler baseline, we also optimize its value to ensure the best possible baseline performance. Specifically, $Z_{max}$ is varied between 0.7 and 1.0 in increments of 0.05, and the configuration yielding the lowest overall RMSE is recorded as the baseline's performance.

## *4.2. Performance Evaluation on the Synthetic Dataset*

### *4.2.1. Dataset Overview and Model Setup*

To evaluate the model's performance under controlled and fully-known conditions, we used the End-Use Load Profile (EULP) dataset [45]. This dataset, developed by the U.S. Department of Energy, provides OpenStudio building energy simulation models and simulated results for various buildings that are calibrated by real-world building conditions. Specifically, we selected the ComStock subset, which focuses on commercial building stock in the U.S.A. From the ComStock subset, we sampled five small or medium-sized office buildings (without data centers) from each of three regions: West, South, and North. In total, 15 buildings were selected.

We use weather files from three cities for the three respective regions: West (San Francisco), South (Phoenix), and North (Chicago). These buildings have various floor area, heights, layouts, alongside various installed lighting and plug capacity, as well as various HVAC capacity and system setups. However, in the original building energy simulation models, fractional schedules for occupancy, lighting, and plug loads follow simple deterministic schedules that remain unchanged across all days.

To create more realistic occupancy profiles, we followed the workflow presented in [40] and utilized a widely-used online occupancy simulator [46]. This simulator generated stochastic daily occupancy profiles, which were then used to derive lighting and plug fractional schedules, as summarized in Table 1. The simulated results span over one year, we take the first half as the train set and the second half as the test set. Figure 7a illustrates the generated stochastic occupancy profiles, demonstrating satisfactory variability. As introduced in Section 4.1.2, to compare discrete occupancy inference performances, the continuous occupancy ratios were categorized into three discrete levels: low, medium, and high, as shown in Figure 7b. Since we configured the simulation setup ourselves, we had prior knowledge of typical occupancy patterns, such as the main occupied period typically occurring between 9h and 18h. Additionally, the simulator's occupant group schedule preferences introduced dispersed patterns with lower peak occupancies and shorter high-occupancy durations.

Based on this knowledge, we used the ASHRAE office schedule as the base reference schedule but made the following adjustments:
- Occupancy ratios outside the 9h-18h period were set to very low values.
- High-occupancy periods (e.g., 9h–11:h AM and 13h–16h) were modified to include more ramping and dropping patterns rather than maintaining constant high occupancy levels, as shown in Figure 7a.

The default level centroids for occupancy ratios were set to [0, 1/3, 2/3, 1], as described in Section 3.2.1.

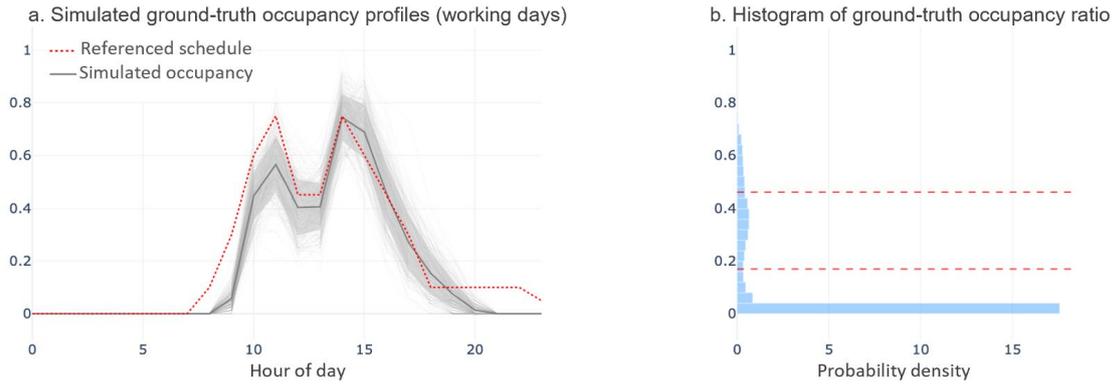

Figure 8 Overview of the simulated stochastic occupancy profiles. (a). Generated stochastic daily occupancy profiles in working days, with the working-day reference schedule in our model highlighted. (b). Histogram of the hourly occupancy ratio, with the bins to categorize the three discrete occupancy levels highlighted.

Table 1 Summary of the EULP synthetic building model and stochastic occupancy & occupant-driven load fractional schedules.

| Dataset | | Schedule simulation | |
|---|---|---|---|
| Name | EULP dataset | Occupants | Two occupant groups: regular: 8.30 am-6.15 pm (variation: 30 mins); flexible: 9.30 am-5 pm (variation: 60 mins); Lunch at 12 am (40 mins), non-working day 5% of max. occupancy |
| Number of buildings | Sample 5 small/med. size office buildings in each West, South, North region in the USA, 15 in total | | |
| Area | 697 - 3484 m² | Lighting | On/off in each room. On when at least one occupant is in the zone. 15-min delay off when no occupant is present |
| Time step | Hourly | | |
| Time span | One year | | |
| Energy metering aggregation level scenarios | Separate: lighting + plug  Lumped: lighting + plug + HVAC | Plug | Energy use varies proportionally with occupant counts |
| Weather files | TMY weather file for each region: West - San Francisco, South - Phoenix, North - Chicago | HVAC | Original setpoint and operation schedules in each building model |

Additionally, the ComStock dataset provided essential metadata for each building, such as construction year and floor area. It also included typical installed capacities for lighting and plug loads per unit area, segmented by construction intervals. These metadata enabled us to initialize the occupant-driven load module parameters (e.g., total lighting and plug load capacities) with rich prior knowledge.

For the "separate" scenario, only the occupant-driven load module was included in the load disaggregator, as HVAC loads were metered separately. For the "lumped" scenario, both occupant-driven and weather-driven load modules were incorporated into the load disaggregator to handle the combined loads. The outdoor air temperature from the weather data is set as the exogenous input to the weather-driven load module. For the KAN spline layer, we set parabolic (2nd order) splines at the normalized range [-2, 2] with 5 grids.

*4.2.2. Occupancy Inference Performance Benchmark ("Separate" Scenario)*

In the "separate" scenario, our model demonstrates high accuracy in inferring discrete occupancy levels. As shown in Table 2, the model achieved a macro average F1 score of approximately 0.8 and a weighted F1 score of about 0.9. All baseline methods also performed reasonably well, which is unsurprising given the relatively simple simulation setup for lighting and plug fractional schedules. Similar trends can also be observed for the continuous occupancy ratio inference, although still falling behind our

OccuEMBED. Even the simple linear baseline can achieve the overall RMSE less than 0.15. Figure 8a gives an example of the accurate inference results from OccuEMBED and the linear scaler.

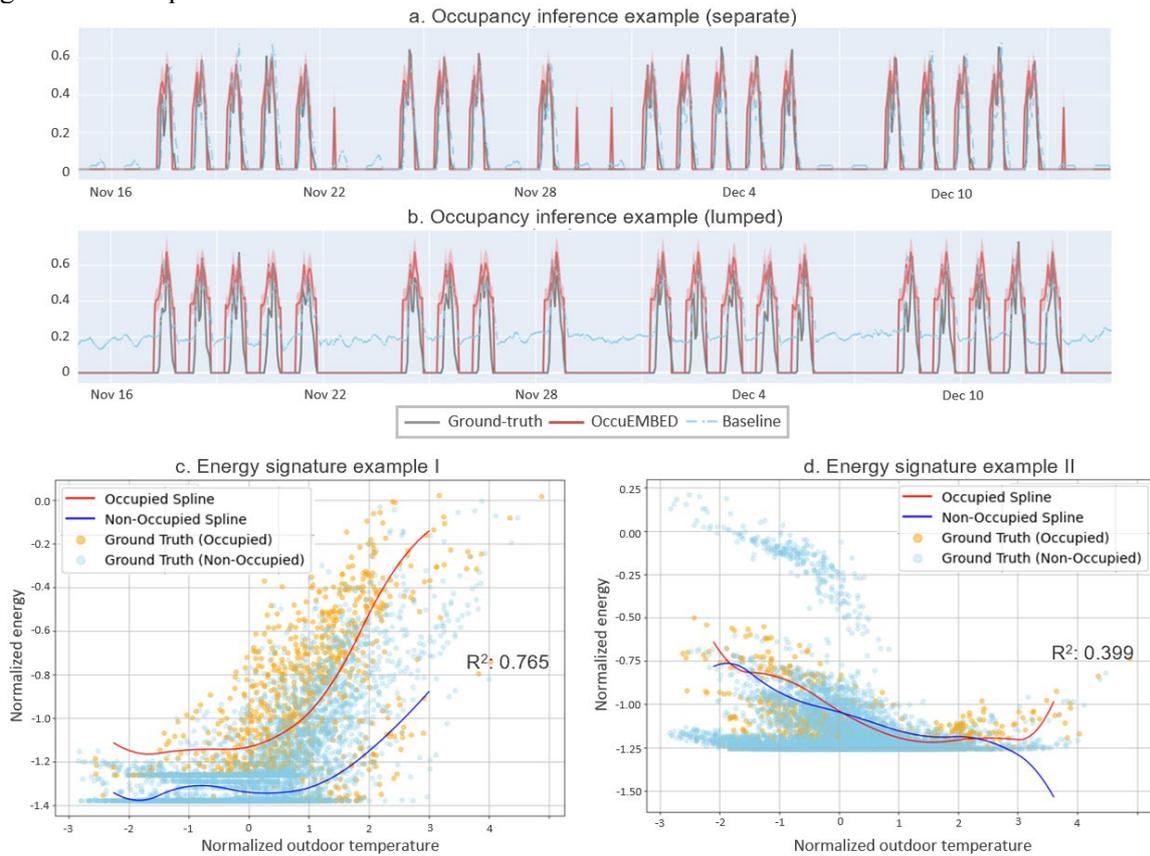

Figure 9 Results of the synthetic dataset performance evaluation. (a). Example of continuous occupancy inference sequence (1: "separate" scenario. 2: "lumped" scenario), with results of the linear scaler baseline for comparison. (b). ES plots of the HVAC load fitted in the KAN spline layers with $R^2$ labelled (1: only electrical cooling. 2: both electric heating and cooling scenario).

Table 2 Performance benchmark on the synthetic dataset "separate" scenario (above: F1 score for discrete levels; below: RMSE for continuous ratios).

| Discrete occupancy levels- F1 score | | | | | |
|---|---|---|---|---|---|
| | Low | Med. | High | Macro Avg. | Weighted Avg. |
| OccuEMBED | 0.944±0.015 | 0.666±0.126 | 0.730±0.142 | 0.780±0.085 | 0.891±0.029 |
| K-means | 0.946±0.013 | 0.199±0.154 | 0.460±0.045 | 0.535±0.056 | 0.803±0.031 |
| GMM | 0.937±0.007 | 0.143±0.047 | 0.466±0.042 | 0.515±0.018 | 0.789±0.016 |
| HMM | 0.945±0.011 | 0.234±0.157 | 0.451±0.070 | 0.543±0.055 | 0.806±0.029 |
| | | | | | |
| Continuous occupancy ratios - RMSE | | | | | |
| | Low | Med. | High | Overall | |
| OccuEMBED | 0.204±0.034 | 0.122±0.025 | 0.101±0.054 | 0.105±0.022 | |
| Linear scaler | 0.173±0.069 | 0.177±0.042 | 0.227±0.129 | 0.127±0.037 | |

Examining individual building performance in Figure 9a and b, we can observe our OccuEMBED consistently delivered strong results, with macro avg. F1 score above 0.7 and RMSE less than 0.12, with the exception of a few outliers. Upon further investigation, these outliers were found to be very small office buildings, which displayed some distinct occupancy patterns, with longer peaking occupancy hours and higher peak occupancy ratio despite the same simulation setup. The reference schedules were not aligned well with these buildings and led to poorer performances.

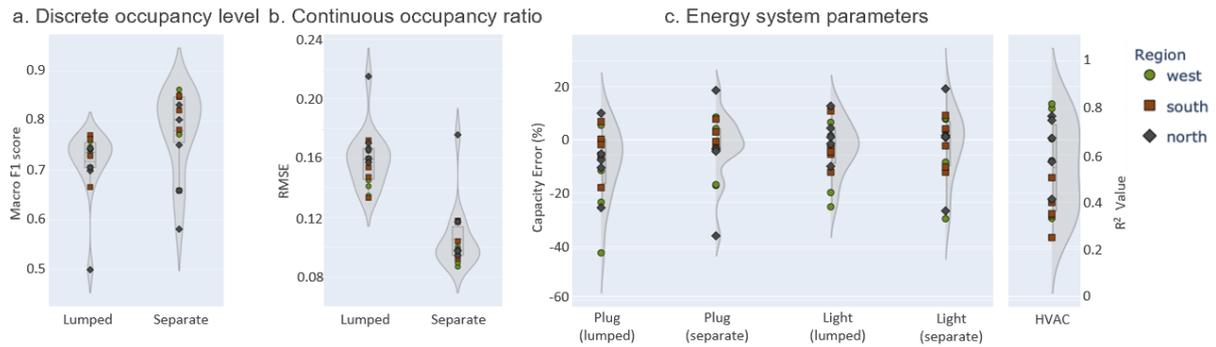

Figure 10 Violin charts of the OccuEMBED performance on each individual building. (a). Macro avg. F1 score of discrete levels. (b). Overall RMSE for the continuous ratios. (c). Percentage error for the occupant-driven load capacities and $R^2$ for the weather-driven HVAC ES pattern.

*4.2.3. Occupancy Inference Performance Benchmark ("Lumped" Scenario)*

While the "lumped" scenario introduces added complexity through aggregating weather-driven loads, our model remained robust and continued to perform well under these more challenging conditions. As shown in Table 3, the macro and weighted F1 scores only experienced a moderate decrease of around 0.1, and RMSE increased slightly by about 0.05. Figure 9 provides a more detailed comparison of performances between the "separate" and "lumped" scenarios across individual cases

Figure 11a and b provide a representative example of how the nuances between the two scenarios manifest in the inferred occupancy profiles for the same building. In the "separate" scenario, the model's inferred occupancy closely mirrored the ground truth. In the more complex "lumped" scenario, it continued to capture the overall patterns and peak levels accurately, though some deviations emerged during morning and afternoon transition periods—where occupancy-driven and weather-driven signals are most entangled. These deviations reflect the inherent challenge of distinguishing overlapping system-level influences in transitional periods. Further examination by occupancy level showed that the performance drop was mainly concentrated in the low and medium ranges, consistent with the increased uncertainty in transition periods. Importantly, even under these more challenging conditions, our model consistently outperformed all baseline methods, which exhibited even greater performance declines.

Table 3 Performance benchmark on the "lumped" scenario of the synthetic dataset (above: F1 score for discrete levels; below: RMSE for continuous ratios).

| Discrete occupancy levels- F1 score | | | | | |
|---|---|---|---|---|---|
|  | Low | Med. | High | Macro Avg. | Weighted Avg. |
| OccuEMBED | 0.873±0.016 | 0.534±0.088 | 0.746±0.114 | 0.717±0.067 | 0.816±0.023 |
| K-means | 0.873±0.104 | 0.392±0.168 | 0.371±0.157 | 0.545±0.096 | 0.768±0.099 |
| GMM | 0.814±0.182 | 0.167±0.148 | 0.392±0.116 | 0.458±0.115 | 0.689±0.156 |
| HMM | 0.904±0.101 | 0.456±0.195 | 0.369±0.130 | 0.576±0.101 | 0.802±0.100 |

| Continuous occupancy ratios - RMSE | | | | |
|---|---|---|---|---|
|  | Low | Med. | High | Overall |
| OccuEMBED | 0.290±0.016 | 0.164±0.013 | 0.079±0.044 | 0.159±0.020 |
| Linear scaler | 0.266±0.095 | 0.187±0.051 | 0.159±0.108 | 0.215±0.041 |

*4.2.4. System-level Load Properties*

For the dynamic occupant-driven load capacities (lighting and plug loads), the percentage error usually fell within the ±20% range, as Figure 9c shows. Notably, the performance decline in the lumped scenario was not significant either. This accuracy is probably attributed to the plausible initial parameter values provided by the occupant-centric load modules, as explained in Section 4.2.1. The outliers with large error also primarily consisted of very small office buildings, as discrepancies in inferred occupancy ratios can result in inaccuracies when estimating occupant-driven loads. The model compensated for these errors by adjusting the total occupant-driven load parameters, leading to further deviations. In addition, since smaller buildings inherently have lower absolute installed capacities, even small absolute errors in load estimation were amplified in percentage errors.

The $R^2$ of the HVAC load ES fitting exhibited a wider and more dispersed distribution, ranging from less than 0.4 to approximately 0.8, as shown in Figure 9c. This can be explained by the two predominant types of ES patterns. The first type, resembling Figure 8c, primarily reflects electricity consumption for cooling, with minimal heating-related electricity usage. This type appears mainly in the buildings in the west region, where heating demand is low due to the moderate climate. It also appears in several buildings in the cold North region, where the heating system is not electrified and the low temperature range exhibits limited electricity consumption of auxiliary systems. The other type, resembling Figure 8d, reflects both electrified cooling and heating demands. This pattern is more common in the south region, where cold and heat exist. The simulated demand in these buildings tends to fluctuate rapidly, introducing significant variations. Such large variations often result in bifurcations and scattering in the ES plots. Since the spline layer used in the model captures only static conditions, it struggles to fit these complex patterns effectively. As a result, in Figure 9c, most of the buildings with low $R^2$ in are from the south region.

## 4.3. Performance Evaluation on Real-World Datasets

*4.3.1. Dataset Overview and Model Setup*

To evaluate the model's performance in real-world conditions, we conducted performance evaluation on three datasets with diverse metering configurations and data availability:
1. Dataset from the study Self-Supervised Indoor Occupancy Estimation for Intelligent Building Management (SSIOE) [12]: this dataset involves a large office building with load metering data available for each floor. Details regarding system-level submetering are not explicitly provided in the dataset. Also, weather data is not included in the original dataset, either. We therefore acquired historical weather data for the building's location and time span using open historical weather data. However, the inclusion of weather data appeared unnecessary as the load metering data showed no correlation with outdoor temperature, suggesting that it primarily consists of occupant-centric loads.
2. LBNL-Bldg 59 Dataset [47]: this dataset comes from Building 59 at the Lawrence Berkeley National Laboratory (LBNL) and includes a comprehensive array of data, including: system-level energy metering, occupant counting, HVAC operation data, and indoor and outdoor environmental conditions. The dataset spans over three years, and we focused on the pre-COVID period, which features complete occupant counts. The detailed system-level submetering allows us to evaluate both the separate and lumped metering scenarios, as well as the model's ability to extract system-level insights.
3. Benchmark8760 Dataset [48]: the Benchmark8760 platform provides historical load data for a dozen large office buildings, along with whole-building occupant counts. Since the time span and data quality vary across buildings, we selected four buildings with complete datasets spanning at least six months. This dataset only includes whole-building lumped load metering for each case, which makes it suitable for testing the model under the lumped metering scenario.

Table 4 gives an overview of the collected real-world datasets. Since these datasets provide occupant counts rather than occupancy ratios, we normalized the counts to derive occupancy ratios. The occupant count distributions exhibit long-tail characteristics due to occasional visitors beyond the regular occupant group. To account for this, we used the 99.9% quantile as the normalization factor instead of the maximum value. Consistent with the processing of synthetic dataset (Section 4.2.1), we also categorized ground-truth occupancy into three levels to evaluate discrete occupancy inference. All the datasets were down-sampled to an hourly step. Depending on the data length for each building, we designated the last one to three months of data as the test set.

Given the limited prior knowledge of occupancy patterns in these real-world datasets, we used the default ASHRAE office occupancy schedule as the reference, as already shown in Figure 4. For the initial lighting and plug capacities, referring to the ASHRAE Standard 90.1 [49], we set both intensities as 8 W/m$^2$ as a general assumption. All other model configurations remained consistent with the setup described in Section 4.2.1.

Table 4 Summary of each real building dataset condition.

| Name | SSIOE dataset | LBNL-Bldg 59 | Benchmark 8760 |
| --- | --- | --- | --- |
| Location | Taipei, Taiwan, China | Berkeley, California, USA | New York, New York, USA |
| Number of buildings | 1 | 1 | 4 (Bldg C, D, F, I) |
| Area | c.a. 37,500 m$^2$ | c.a. 1600 m$^2$ | 14,694 - 176,700 m$^2$ |
| Time step | Hourly | Hourly | Hourly |
| Time span | 06/2018-01/2019 | 05/2018 - 02/2019 | max. 06/2022-04/2023 |
| Energy metering aggregation level scenarios | Separate: obviously no HVAC | Separate: lighting + plug  Lumped: lighting + plug + HVAC | Lumped: lighting + plug + HVAC |
| Weather data | Weather data unnecessary | Site measurements available | Local weather data collected in the data platform |

### 4.3.2. Performance Evaluation on the SSIOE Dataset

The SSIOE dataset contains only occupant-driven loads, and these loads appear to follow occupancy levels almost perfectly. Consequently, as Table 5 shows, while our model still outperformed all the clustering baselines, the performance differences were minimal across all evaluation metrics.

Interestingly, in continuous occupancy ratio inference, the simple linear scaler baseline outperformed our model. Figure 10 provides insights into this aspect. While our model correctly identified peak occupancy levels (achieving lower RMSE than the linear scaler in the high occupancy range) and captured the overall patterns well, it exhibited noticeable mismatches during unoccupied periods. These mismatches were reflected in significantly higher RMSE values for the low and medium occupancy levels compared to the linear scaler.

This result highlights a limitation of the current model: the generated occupancy profile samples for inactive periods lack sufficient variability, making it difficult to identify accurate occupancy patterns during these times. Addressing this issue may require further incorporating additional prior knowledge about typical occupancy patterns during unoccupied periods.

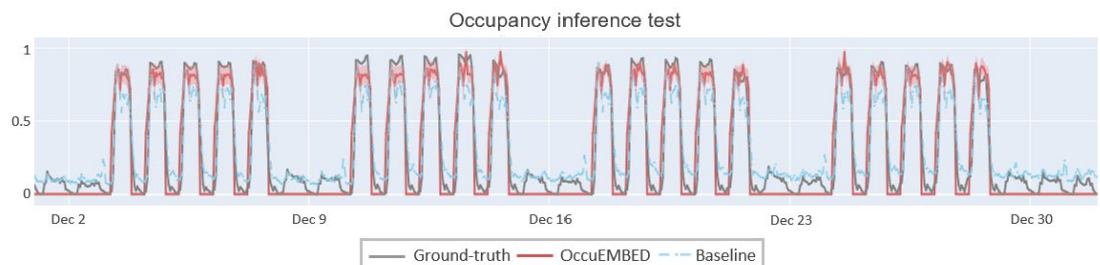

Figure 12 Example of occupancy inference sequence on the SSIOE dataset

Table 5 Performance benchmark on the SSIOE dataset (above: F1 score for discrete levels; below: RMSE for continuous ratios).

| Discrete occupancy levels- F1 score | | | | | |
|---|---|---|---|---|---|
| | Low | Med. | High | Macro Avg. | Weighted Avg. |
| OccuEMBED | 0.946 | 0.563 | 0.982 | 0.83 | 0.929 |
| K-means | 0.953 | 0.515 | 0.859 | 0.776 | 0.898 |
| GMM | 0.882 | 0.364 | 0.919 | 0.722 | 0.856 |
| HMM | 0.902 | 0.345 | 0.948 | 0.732 | 0.876 |

| Continuous occupancy ratios - RMSE | | | | |
|---|---|---|---|---|
| | Low | Med. | High | Overall |
| OccuEMBED | 0.138 | 0.255 | 0.079 | 0.137 |
| Linear scaler | 0.121 | 0.204 | 0.096 | 0.122 |

### 4.3.3. Performance Evaluation on the LBNL-Bldg 59 Dataset

As an office building in the U.S., the LBNL-Bldg 59 dataset offers a unique opportunity to compare synthetic EULP building models (Section 4.2) with real-world complexities.

In the "separate" scenario, all models exhibited reduced performance compared to their results on synthetic data, as shown in Table 6. Both discrete occupancy level inference and continuous ratio inference were compromised across all models. This suggests that, in real-world small office buildings, occupant-driven loads (e.g., lighting and plugs) do not strictly follow occupancy ratios and instead display more variability. Despite this added complexity, our model maintained a clear lead over the baseline methods in both discrete and continuous metrics.

Table 6 Performance benchmark on the "separate" scenario of the LBNL-Bldg 59 dataset (above: F1 score for discrete levels; below: RMSE for continuous ratios).

| Discrete occupancy levels- F1 score | | | | | |
|---|---|---|---|---|---|
| | Low | Med. | High | Macro Avg. | Weighted Avg. |
| OccuEMBED | 0.898 | 0.591 | 0.456 | 0.648 | 0.8 |
| K-means | 0.933 | 0.247 | 0.449 | 0.543 | 0.751 |
| GMM | 0.887 | 0.363 | 0.504 | 0.585 | 0.748 |
| HMM | 0.908 | 0.257 | 0.453 | 0.539 | 0.736 |

| Continuous occupancy ratios - RMSE | | | | |
|---|---|---|---|---|
| | Low | Med. | High | Overall |
| OccuEMBED | 0.121 | 0.229 | 0.131 | 0.152 |
| Linear scaler | 0.153 | 0.32 | 0.097 | 0.198 |

In the "lumped" scenario, as Table 7 shows, the clustering-based baselines experienced significant drops in F1 metrics, while our model's remained largely unaffected. That said, when examining the RMSE at each occupancy level, we observed a noticeable increase in errors for the low occupancy range, which impacted the overall RMSE. Figure 11a and b further illustrate how the added weather-driven loads influenced the inferred occupancy profiles. Similar to the case of synthetic dataset (Section 4.2.3), the weather-driven variability led to an overall increase in inferred occupancy ratios and inaccuracies particularly during transition periods in the morning and afternoon hours.

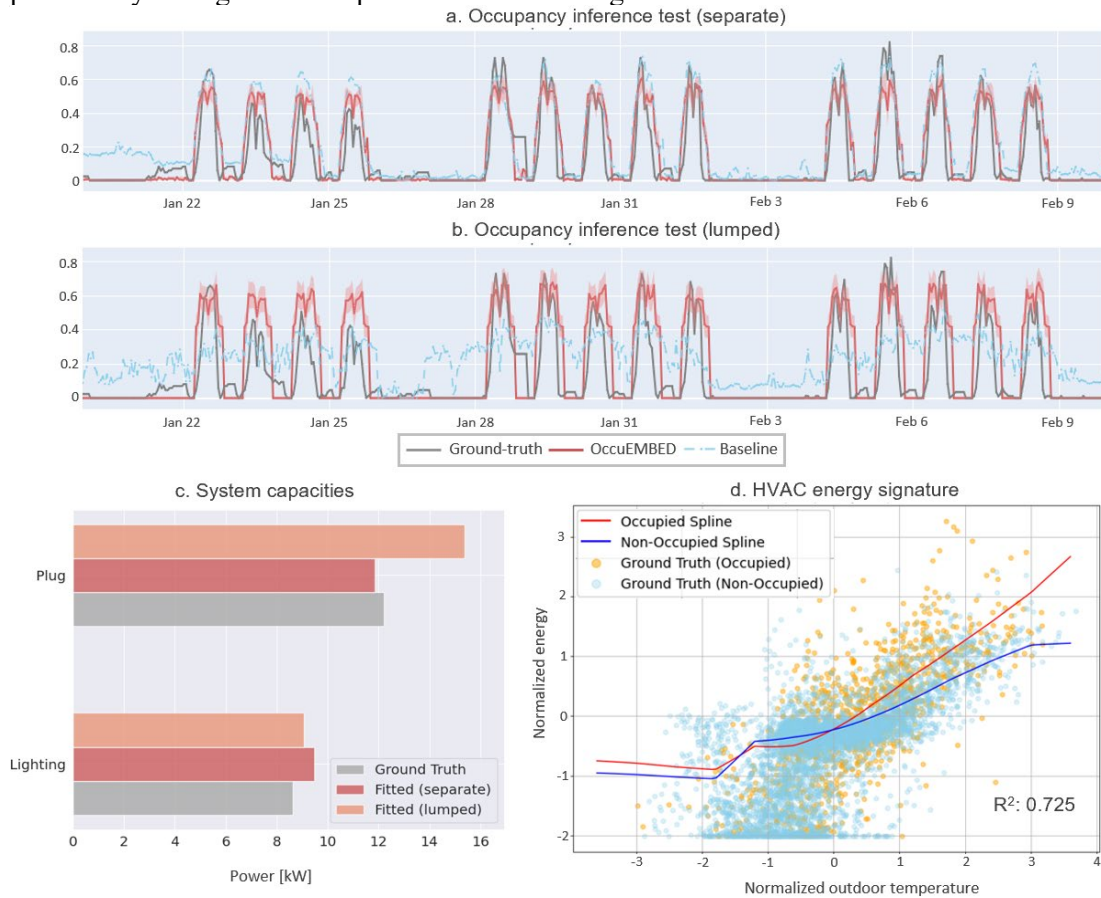

Figure 13 Results of the LBNL-Bldg 59 dataset performance evaluation. (a). Example of continuous occupancy inference sequence (1: "separate" scenario. 2: "lumped" scenario), with results of the linear scaler baseline for comparison. (b). Evaluation of the occupant-driven system capacities. (c). ES plot of the HVAC load fitted in the KAN spline layers with $R^2$ labelled.

Table 7 Performance benchmark on the "lumped" scenario of the LBNL-Bldg 59 dataset (above: F1 score for discrete levels; below: RMSE for continuous ratios).

| Discrete occupancy levels- F1 score | | | | | |
|---|---|---|---|---|---|
| | Low | Med. | High | Macro Avg. | Weighted Avg. |
| OccuEMBED | 0.888 | 0.566 | 0.452 | 0.635 | 0.795 |
| K-means | 0.801 | 0.262 | 0.182 | 0.415 | 0.652 |
| GMM | 0.801 | 0.259 | 0.168 | 0.409 | 0.651 |
| HMM | 0.82 | 0.205 | 0.218 | 0.414 | 0.657 |
| | | | | | |
| Continuous occupancy ratios - RMSE | | | | | |
| | Low | Med. | High | Overall | |
| OccuEMBED | 0.208 | 0.227 | 0.129 | 0.181 | |
| Linear scaler | 0.215 | 0.223 | 0.364 | 0.222 | |

The detailed submetering in this dataset allowed us to evaluate the load disaggregator parameters more thoroughly. For the occupant-driven systems (lighting and plugs), we lacked direct ground truth for dynamic installed capacities, unlike in the synthetic dataset. Instead, we estimated these capacities by selecting the maximum observed loads for lighting and plugs and subtracting a low quantile (representing base load levels). As shown in Figure 11c, our model accurately identified the capacities for both systems, even under the lumped scenario, with minimal absolute error. Our model also exhibits effective fitting of HVAC load Energy Signature (ES) patterns, as illustrated in Figure 11d. This building does not seem to have setback or switch-off schedules during unoccupied periods. The slight discrepancies in HVAC loads between occupied and unoccupied periods were likely due to: higher internal gains during occupied periods, and potential changes in temperature setpoints made by occupants. The two KAN spline layers for occupied and unoccupied periods respectively, successfully captured these discrepancies. However, the transition temperature range, where displays low values and significant variability in HVAC loads, was still not fully captured by the model.

### 4.3.4. Performance Evaluation on the Benchmark 8760 Dataset

The Benchmark 8760 dataset only provides whole-building lumped metering data in all four buildings evaluated. As Table 8 shows, our model consistently outperformed the baseline methods. However, we observed performance compromises in all models comparable to those seen in other "lumped" scenarios.

Table 8 Performance benchmark on each building in the Benchmark 8760 dataset (above: F1 score for discrete levels; below: RMSE for continuous ratios).

| Discrete occupancy levels - F1 score | | | | | | | | |
|---|---|---|---|---|---|---|---|---|
| | Bldg C - Avg. | | Bldg D - Avg. | | Bldg F - Avg. | | Bldg I - Avg. | |
| | Macro | Weighted | Macro | Weighted | Macro | Weighted | Macro | Weighted |
| OccuEMBED | 0.575 | 0.759 | 0.686 | 0.825 | 0.617 | 0.772 | 0.651 | 0.798 |
| K-means | 0.425 | 0.662 | 0.436 | 0.7 | 0.478 | 0.636 | 0.435 | 0.651 |
| GMM | 0.464 | 0.696 | 0.485 | 0.682 | 0.372 | 0.635 | 0.473 | 0.688 |
| HMM | 0.434 | 0.681 | 0.437 | 0.726 | 0.42 | 0.59 | 0.546 | 0.71 |

| Continuous occupancy ratios - RMSE | | | | |
|---|---|---|---|---|
| | Bldg C - overall | Bldg D - overall | Bldg F - overall | Bldg I - overall |
| OccuEMBED | 0.214 | 0.144 | 0.215 | 0.178 |
| Linear scaler | 0.267 | 0.282 | 0.297 | 0.266 |

As shown in Figure 12a and b, the inferred occupancy profiles seem to lack sufficient variability to fully capture the true occupancy dynamics. This issue may stem from several factors: First, these large office buildings with significant occupant counts likely have high capacities for centrally controlled lighting and substantial base-load appliances. These factors, combined with the influence of weather-driven loads, reduce the prominence of occupant-driven variations in the lumped metering data. Consequently, the inferred occupancy profiles often appeared similar across days with limited day-to-day nuances. Additionally, the limited variability in generated candidate occupancy samples exacerbates this issue, especially for low-occupancy periods such as non-working days or working days that are still subject to holiday effects. For these days, our model aimed to match the trends using the closest available occupancy profiles, but mismatches were inevitable due to insufficient sample diversity.

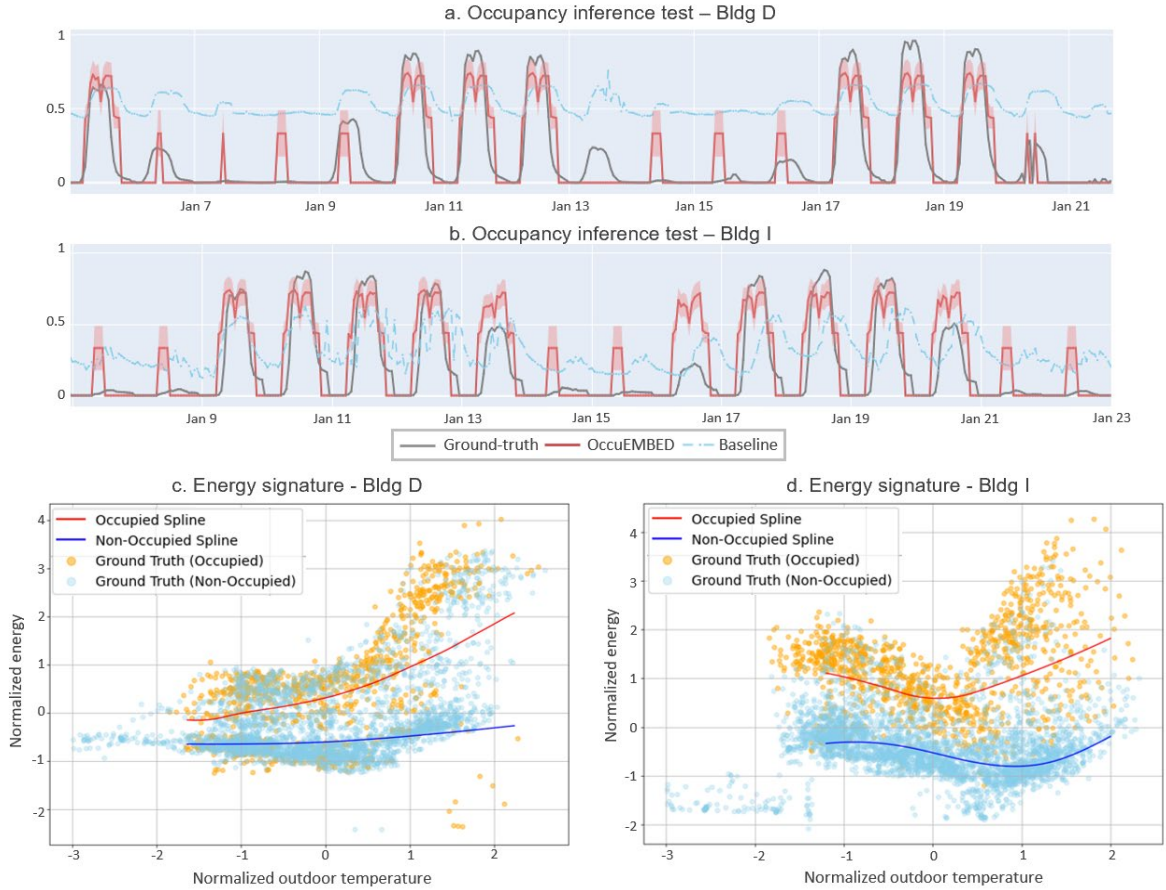

Figure 14 Results of the Benchmark 8760 dataset performance evaluation. (a). Example of continuous occupancy inference sequence (1: Bldg D. 2: Bldg I), with results of the linear scaler baseline for comparison. (b). ES plots on the whole-building load fitted in the KAN spline layers with $R^2$ labelled and adjusted offsets (1: Bldg D. 2: Bldg I).

Still, as shown in Figure 12b and c, the KAN spline layers effectively captured weather-driven trends in the lumped loads. Since HVAC sub-metering is not available, we plot our splines together with the ES of the lumped metering. Here, we adjusted their offsets in splines based on the average occupant-driven load during occupied and unoccupied periods to align them together. Here, buildings C, F, and D followed a pattern with electricity usage primarily driven by cooling demand and clear operational setbacks during unoccupied periods, as shown in Figure 12c. The occupied and unoccupied splines in OccuEMBED successfully captured this operational behavior. Building I exhibited another unique pattern where both cooling and heating demands were electrically driven, and no clear setbacks or switch-offs were observed during unoccupied periods, as shown in Figure 12d. Despite this more complex behavior, our model accurately captured the combined heating and cooling patterns as long as no significant bifurcations (such as those observed in the synthetic dataset) were present.

These identified ES patterns highlight the operational characteristics of the buildings and can be used to develop more occupant-centric operation strategies, which will be further discussed below.

## 5. RESULTS: MODEL APPLICATIONS

In this section, we present how the proposed OccuEMBED can be deployed in real-world scenarios using a building load monitoring platform. This deployment highlights the model's potential to uncover detailed occupancy patterns, provide system-level insights, and enable occupant-centric operations at scale.

## 5.1. Occupancy Inference and System-level Insights from Historical Building Loads

The proposed model can offer powerful applications, especially for utility providers and large building portfolio managers, who monitor and manage energy usage across multiple office buildings. Figure 13 illustrates how the model can seamlessly integrate with existing building load monitoring platforms, providing real-time inference and historical analysis of underlying occupancy ratios. Since the model requires only historical load data as input, it is both easy to deploy and highly scalable. In addition to occupancy inference, the model generates system-level insights by breaking down energy consumption and identifying power capacities by system type. The inferred average occupancy profiles can also be reviewed by experienced managers to validate patterns and make adjustments.

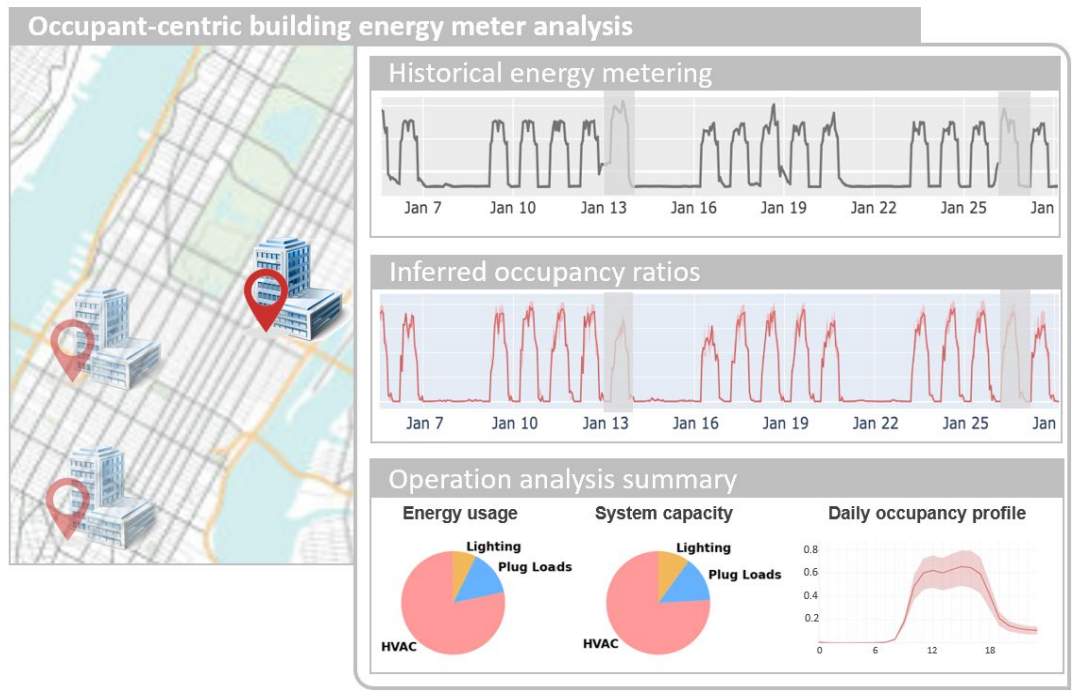

Figure 15 Illustration of how the OccuEMBED can be integrated with a building load monitoring platform to provide occupancy inference and system-level analysis, using data from Bldg F in the Benchmark 8760 dataset. Here, highlighted periods in the historical loads and occupancy inference profiles indicating high loads mismatched with low occupancy.

Besides understanding occupancy, the model enables actionable insights to promote occupant-centric operations. For instance, Figure 13 highlights periods where high energy loads coincide with low occupancy, likely due to lighting or HVAC systems operating regardless of occupancy levels. Identifying such mismatches allows utilities and managers to adjust operation schedules or occupant spatial allocation, fostering more personalized and responsive energy system operations. These changes can reduce overall energy usage and provide greater flexibility for downward demand response. Although assessing such complex "what-if" operational scenarios beyond the scope of our model, it is indeed capable of assessing some simpler adjustments for more occupant-centric operations. Below we demonstrate how occupant-responsive HVAC operation strategies can be evaluated by leveraging the learned ES patterns captured in the KAN spline layers.

## 5.2. "What-if" Assessment for more Occupant-responsive Operation Strategies

As discussed in Section 4.3.4, for Bldg C, D, and F, clear evidence of HVAC system switch-off during unoccupied periods was observed, as illustrated in Figure 14. However, closer examination revealed that high load values persisted during some nominally unoccupied periods (8 pm to midnight), likely indicating active HVAC operation. This discrepancy suggests that the existing switch-off schedules were not properly aligned with actual occupancy patterns. To assess a more occupant-responsive switch-off strategy, we adjusted the loads by subtracting the trend from the occupied spline for this interval and replacing it with the trend from the unoccupied spline. In Bldg F, this adjustment reduced the total building energy consumption from 301 MWh to 289 MWh.

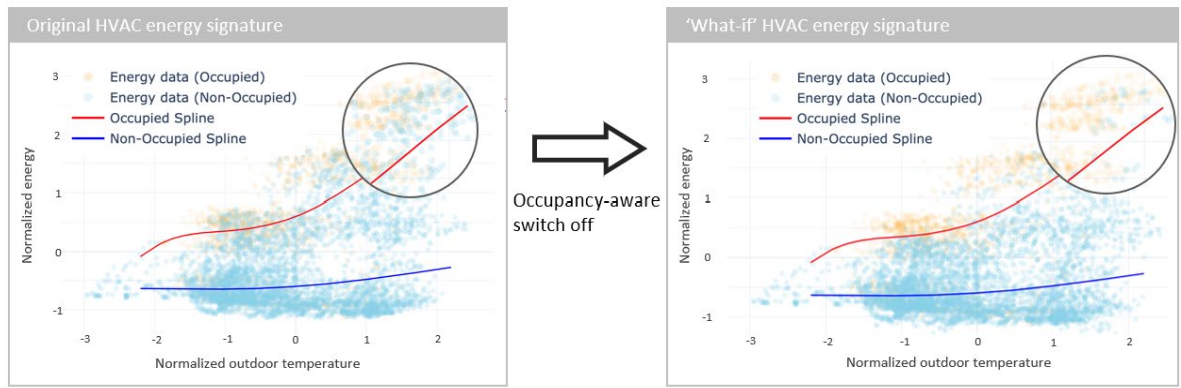

Figure 16 "What-if" assessment on adjusting on-off HVAC schedule according to the actual unoccupied period for the Bldg F in the Benchmark 8760 dataset (occupied/unoccupied period differentiation using inferred occupancy). Comparison of the before and after ES plots.

## 6. DICSUSSION

### 6.1. Advantages of the Model Performance and Prospect of Application

Our primary contribution with OccuEMBED is an integrated framework that effectively combines energy analysis and occupancy inference, requiring only building-level energy metering as input. This minimal data requirement ensures practicality and scalability, making the model highly suitable for large-scale deployment across diverse building portfolios.

Performance evaluations across both synthetic and real-world datasets consistently demonstrated our model's improvement over existing baselines. This strong performance stems from two key innovations: the incorporation of prior knowledge to ensure credible occupancy patterns and the explicit representation of occupancy-energy-environment interactions. These design choices allow for highly accurate occupancy inference while maintaining model interpretability and physical consistency, making it applicable to real-world building operations.

While clustering-based baselines performed reasonably well, they lacked an explicit representation of occupancy behaviors and energy system dynamics. Among them, the HMM model, which accounts for heterogeneous occupancy transitions across different times of the day, exhibited relatively robust performance. However, without direct integration of prior knowledge, it could not match the accuracy and adaptability of our approach. The linear scaler baseline, while capable of generating continuous occupancy ratios, struggled in the lumped metering scenario. Though its RMSE metrics seemed to not be significantly affected after the weather trend removal, its output remained clustered around mid-range values, and it failed to capture meaningful diurnal variations in occupancy profiles (Figure 11b and 12b).

Beyond inference accuracy, our model offers practical value in real-world energy management. By extracting both occupancy insights and system-level load characteristics, it provides actionable intelligence for utility providers and large building portfolio managers, enabling more data-driven decision-making for energy optimization. Furthermore, because the model relies only on historical load data, it can be seamlessly integrated into load forecasting frameworks, extending its utility for predicting future occupancy trends and enhancing occupant-centric energy management strategies.

### 6.2. Future Enhancements and Expanding Capabilities

While OccuEMBED achieves strong performance and broad applicability, there are opportunities for further enhancements to increase its flexibility and generalization capabilities.

A key area for improvement is the variability of generated candidate occupancy samples. Currently, the model perturbs a single reference schedule for working days and another for non-working days. However, as discussed in Section 4.3, real-world occupancy schedules can exhibit greater diversity, influenced by factors such as building type, tenant business, and seasonal variations. Incorporating a richer set of reference schedules—including variations for busy, normal, idle, and holiday days—could improve the model's ability to generalize across different operational contexts. Such reference schedule expansions could be informed by site studies in urban energy research, strengthening the model's adaptability.

Another potential improvement lies in the load disaggregator module. While the KAN splines effectively captured static ES patterns, it does not account for pre-cooling, pre-heating, or demand-flexible HVAC strategies. Future work could explore thermodynamics-based alternatives to better model these operational complexities. Additionally, as mentioned in Section 5.2, many buildings operate their central HVAC and lighting systems on fixed schedules rather than occupancy-responsive controls. To accommodate such cases, the load disaggregator could incorporate schedule-driven operation modules, allowing predefined system schedules to override occupancy-driven assumptions where applicable. Implementing this would be straightforward (e.g., modifying non-occupancy and full-occupancy levels based on known schedules) but would require collaborative iteration with users to identify the most representative schedules.

For systems with hybrid controls—such as lighting systems that respond to both occupancy sensors and daylight sensors—a more integrated approach may be beneficial. Future enhancements could explore connecting the occupant-driven and weather-driven modules to better reflect such interdependent system behaviors where necessary.

Overall, OccuEMBED achieves a unique balance of scalability and adaptability. On one hand, it enables efficient, large-scale deployment using generic prior knowledge and minimal data input, making it accessible across a wide range of buildings. On the other hand, its building-specific insights allow for an interactive and iterative process, where utilities and facility managers can refine occupancy assumptions and incorporate case-specific knowledge. These refinements, in turn, contribute to an evolving prior knowledge base in the future, enabling greater accuracy as the model scales to larger building portfolios.

### *6.3. The Role of System Submetering*

An additional insight from our evaluation was the importance of system-level submetering. In buildings where occupant-driven loads are separately metered, even simpler clustering models or linear scalers performed well. This underscores the value of submetering for improving energy analysis and operational insights. However, this observation does not diminish the significance of our model—instead, it reinforces its scalability advantage. By leveraging detailed system-level data from submetered buildings, prior knowledge can be refined in our model, which can then be appliedeffectively to the vast majority of buildings where only lumped metering is available. This transferability is key to achieving broad deployment and ensuring that occupancy-aware energy management strategies can be adopted at scale.

### 7. CONCLUSION

In this study, we propose OccuEMBED, a novel integrated framework for inferring occupancy levels and disaggregating energy loads using only building-level energy metering. Due to the two key designs (i.e., a probabilistic occupancy profile generator and a controllable and interpretable load disaggregator) this model enables the incorporation of prior knowledge from engineering practice into the data-driven workflow in an explicit, interpretable, and interactive manner. The model demonstrated marked improvements in performance across synthetic and real-world datasets. Its adaptivity and minimal data requirements make it a scalable and credible tool for building portfolio managers and utilities. This framework lays a robust foundation in advancing intelligent and occupant-centric building management systems at scale.

## 8. ACKNOWLEDGEMENT

This research was supported by EPFL and the ETH-Domain Joint Initiative program through the UrbanTwin project. We gratefully acknowledge the Benchmark 8760 initiative for sharing their dataset.

**CRediT authorship contribution statement**
**Zhang Yufei**: Conceptualization, Methodology, Data Curation, Software, Formal analysis, Writing - Original Draft. **Andrew Sonta**: Conceptualization, Methodology, Supervision, Writing - Review & Editing, Funding acquisition.

## REFERENCES

[1] "Buildings - Energy System," IEA. Accessed: Dec. 01, 2023. [Online]. Available: https://www.iea.org/energy-system/buildings

[2] N. E. Klepeis et al., "The National Human Activity Pattern Survey (NHAPS): a resource for assessing exposure to environmental pollutants," *J. Expo. Sci. Environ. Epidemiol.*, vol. 11, no. 3, Art. no. 3, Jul. 2001, doi: 10.1038/sj.jea.7500165.

[3] T. Hong and H.-W. Lin, "Occupant Behavior: Impact on Energy Use of Private Offices," Jan. 2012, p. 8.

[4] A. Boodi, K. Beddiar, M. Benamour, Y. Amirat, and M. Benbouzid, "Intelligent Systems for Building Energy and Occupant Comfort Optimization: A State of the Art Review and Recommendations," *Energies*, vol. 11, no. 10, Art. no. 10, Oct. 2018, doi: 10.3390/en11102604.

[5] W. O'Brien et al., "Introducing IEA EBC annex 79: Key challenges and opportunities in the field of occupant-centric building design and operation," *Build. Environ.*, vol. 178, p. 106738, Jul. 2020, doi: 10.1016/j.buildenv.2020.106738.

[6] Z. Nagy et al., "Ten questions concerning occupant-centric control and operations," *Build. Environ.*, vol. 242, p. 110518, Aug. 2023, doi: 10.1016/j.buildenv.2023.110518.

[7] M. Hu, F. Xiao, and S. Wang, "Neighborhood-level coordination and negotiation techniques for managing demand-side flexibility in residential microgrids," *Renew. Sustain. Energy Rev.*, vol. 135, p. 110248, Jan. 2021, doi: 10.1016/j.rser.2020.110248.

[8] R. Melfi, B. Rosenblum, B. Nordman, and K. Christensen, "Measuring building occupancy using existing network infrastructure," in *2011 International Green Computing Conference and Workshops*, Jul. 2011, pp. 1–8. doi: 10.1109/IGCC.2011.6008560.

[9] S. Naylor, M. Gillott, and T. Lau, "A review of occupant-centric building control strategies to reduce building energy use," *Renew. Sustain. Energy Rev.*, vol. 96, pp. 1–10, Nov. 2018, doi: 10.1016/j.rser.2018.07.019.

[10] A. Ebadat, G. Bottegal, D. Varagnolo, B. Wahlberg, and K. H. Johansson, "Estimation of building occupancy levels through environmental signals deconvolution," in *Proceedings of the 5th ACM Workshop on Embedded Systems For Energy-Efficient Buildings*, in BuildSys'13. New York, NY, USA: Association for Computing Machinery, Nov. 2013, pp. 1–8. doi: 10.1145/2528282.2528290.

[11] H. Li, T. Hong, and M. Sofos, "An inverse approach to solving zone air infiltration rate and people count using indoor environmental sensor data," *Energy Build.*, vol. 198, pp. 228–242, Sep. 2019, doi: 10.1016/j.enbuild.2019.06.008.

[12] S.-H. Huang, T.-Y. Chao, B. A. Wibisono, M. P.-H. Lin, and C.-C. Huang, "SSIOE: Self-Supervised Indoor Occupancy Estimation for Intelligent Building Management," *IEEE Trans. Autom. Sci. Eng.*, vol. 21, no. 3, pp. 3025–3038, Jul. 2024, doi: 10.1109/TASE.2023.3273151.

[13] E. Barbour, C. C. Davila, S. Gupta, C. Reinhart, J. Kaur, and M. C. González, "Planning for sustainable cities by estimating building occupancy with mobile phones," *Nat. Commun.*, vol. 10, no. 1, Art. no. 1, Aug. 2019, doi: 10.1038/s41467-019-11685-w.

[14] Z. Wang, T. Hong, M. A. Piette, and M. Pritoni, "Inferring occupant counts from Wi-Fi data in buildings through machine learning," *Build. Environ.*, vol. 158, pp. 281–294, Jul. 2019, doi: 10.1016/j.buildenv.2019.05.015.

[15] Z. Li and B. Dong, "Short term predictions of occupancy in commercial buildings—Performance analysis for stochastic models and machine learning approaches," *Energy Build.*, vol. 158, pp. 268–281, Jan. 2018, doi: 10.1016/j.enbuild.2017.09.052.

[16] Y. Peng, A. Rysanek, Z. Nagy, and A. Schlüter, "Using machine learning techniques for occupancy-prediction-based cooling control in office buildings," *Appl. Energy*, vol. 211, pp. 1343–1358, Feb. 2018, doi: 10.1016/j.apenergy.2017.12.002.

[17] X. Chen, T. Guo, M. Kriegel, and P. Geyer, "A hybrid-model forecasting framework for reducing the building energy performance gap," *Adv. Eng. Inform.*, vol. 52, p. 101627, Apr. 2022, doi: 10.1016/j.aei.2022.101627.

[18] H. Li *et al.*, "Data-Driven Key Performance Indicators and Datasets for Building Energy Flexibility: A Review and Perspectives," Nov. 22, 2022, *arXiv*: arXiv:2211.12252. doi: 10.48550/arXiv.2211.12252.

[19] T. Hong, M. Gui, M. E. Baran, and H. L. Willis, "Modeling and forecasting hourly electric load by multiple linear regression with interactions," in *IEEE PES General Meeting*, Jul. 2010, pp. 1–8. doi: 10.1109/PES.2010.5589959.

[20] M. Mosteiro-Romero, P. Alva, C. C. Miller, and R. Stouffs, "Towards occupant-driven district energy system operation: A digital twin platform for energy resilience and occupant well-being," in *Elements*, 2022. Accessed: Aug. 01, 2023. [Online]. Available: https://scholarbank.nus.edu.sg/handle/10635/243262

[21] A. J. Sonta, P. E. Simmons, and R. K. Jain, "Understanding building occupant activities at scale: An integrated knowledge-based and data-driven approach," *Adv. Eng. Inform.*, vol. 37, pp. 1–13, Aug. 2018, doi: 10.1016/j.aei.2018.04.009.

[22] W. Kleiminger, C. Beckel, T. Staake, and S. Santini, "Occupancy Detection from Electricity Consumption Data," in *Proceedings of the 5th ACM Workshop on Embedded Systems For Energy-Efficient Buildings*, in BuildSys'13. New York, NY, USA: Association for Computing Machinery, Nov. 2013, pp. 1–8. doi: 10.1145/2528282.2528295.

[23] A. Meier and D. Cautley, "Practical limits to the use of non-intrusive load monitoring in commercial buildings," *Energy Build.*, vol. 251, p. 111308, Nov. 2021, doi: 10.1016/j.enbuild.2021.111308.

[24] "Modeling Office Building Occupancy in Hourly Data-Driven and Detailed Energy Simulation Programs - ProQuest." Accessed: Jan. 29, 2025. [Online]. Available: https://www.proquest.com/docview/192558113?pq-origsite=gscholar&fromopenview=true&sourcetype=Scholarly%20Journals

[25] Z. Wang, T. Hong, and M. A. Piette, "Data fusion in predicting internal heat gains for office buildings through a deep learning approach," *Appl. Energy*, vol. 240, pp. 386–398, Apr. 2019, doi: 10.1016/j.apenergy.2019.02.066.

[26] W. O'Brien, I. Gaetani, S. Carlucci, P.-J. Hoes, and J. L. M. Hensen, "On occupant-centric building performance metrics," *Build. Environ.*, vol. 122, pp. 373–385, Sep. 2017, doi: 10.1016/j.buildenv.2017.06.028.

[27] L. G. Housing, "Electricity sub-metering for buildings." Accessed: Apr. 18, 2024. [Online]. Available: https://www.business.qld.gov.au/industries/building-property-development/building-construction/laws-codes-standards/sustainable-housing/electricity-sub-metering

[28] "关于发布浙江省工程建设标准《公共建筑用电分项分区计量系统设计标准》的公告." Accessed: Apr. 18, 2024. [Online]. Available: https://jst.zj.gov.cn/art/2023/4/3/art_1228990170_329.html?eqid=a0a0125c001fc04000000005642cd1f7

[29] "Becoming Smarter about Energy: A Guide to Submeter Development and Greater Energy Management Insights | Better Buildings Initiative." Accessed: Apr. 18, 2024. [Online]. Available: https://betterbuildingssolutioncenter.energy.gov/tools/becoming-smarter-about-energy-a-guide-submeter-development-and-greater-energy-management

[30] B. J. Birt, G. R. Newsham, I. Beausoleil-Morrison, M. M. Armstrong, N. Saldanha, and I. H. Rowlands, "Disaggregating categories of electrical energy end-use from whole-house hourly data," *Energy Build.*, vol. 50, pp. 93–102, Jul. 2012, doi: 10.1016/j.enbuild.2012.03.025.

[31] J. Mathieu, P. Price, S. Kiliccote, and M. Piette, "Quantifying Changes in Building Electricity Use, With Application to Demand Response," *IEEE Trans Smart Grid*, vol. 2, pp. 507–518, Sep. 2011, doi: 10.1109/TSG.2011.2145010.

[32] Nick MacMackin, L. Miller, and R. Carriveau, "Modeling and disaggregating hourly effects of weather on sectoral electricity demand," *Energy*, vol. 188, p. 115956, Dec. 2019, doi: 10.1016/j.energy.2019.115956.

[33] F. Niu, Z. O'Neill, and C. O'Neill, "Data-driven based estimation of HVAC energy consumption using an improved Fourier series decomposition in buildings," *Build. Simul.*, vol. 11, no. 4, pp. 633–645, Aug. 2018, doi: 10.1007/s12273-018-0431-2.

[34] A. Capozzoli, M. S. Piscitelli, S. Brandi, D. Grassi, and G. Chicco, "Automated load pattern learning and anomaly detection for enhancing energy management in smart buildings," *Energy*, vol. 157, pp. 336–352, Aug. 2018, doi: 10.1016/j.energy.2018.05.127.

[35] S. Rouchier, "Bayesian Workflow and Hidden Markov Energy-Signature Model for Measurement and Verification," *Energies*, vol. 15, no. 10, Art. no. 10, Jan. 2022, doi: 10.3390/en15103534.

[36] Y. Zhang and A. Sonta, "OccuVAE: Integrating unsupervised occupancy inference in data-driven energy modeling for human-centric operation," in *Proceedings of the 10th ACM International Conference on Systems for Energy-Efficient Buildings, Cities, and Transportation*, in BuildSys '23. New York, NY, USA: Association for Computing Machinery, Nov. 2023, pp. 398–405. doi: 10.1145/3600100.3626342.

[37] A. Razavi, A. van den Oord, and O. Vinyals, "Generating Diverse High-Fidelity Images with VQ-VAE-2," Jun. 02, 2019, *arXiv*: arXiv:1906.00446. doi: 10.48550/arXiv.1906.00446.

[38] Z. Liu *et al.*, "KAN: Kolmogorov-Arnold Networks," Jun. 16, 2024, *arXiv*: arXiv:2404.19756. doi: 10.48550/arXiv.2404.19756.

[39] D. P. Kroese, R. Y. Rubinstein, and P. W. Glynn, "The Cross-Entropy Method for Estimation," in *Handbook of Statistics*, vol. 31, Elsevier, 2013, pp. 19–34. doi: 10.1016/B978-0-444-53859-8.00002-3.

[40] H. Li, Z. Wang, and T. Hong, "A synthetic building operation dataset," *Sci. Data*, vol. 8, no. 1, Art. no. 1, Aug. 2021, doi: 10.1038/s41597-021-00989-6.

[41] M. Seitzer, A. Tavakoli, D. Antic, and G. Martius, "On the Pitfalls of Heteroscedastic Uncertainty Estimation with Probabilistic Neural Networks," Apr. 01, 2022, *arXiv*: arXiv:2203.09168. doi: 10.48550/arXiv.2203.09168.

[42] A. Paszke *et al.*, "PyTorch: An Imperative Style, High-Performance Deep Learning Library," Dec. 03, 2019, *arXiv*: arXiv:1912.01703. doi: 10.48550/arXiv.1912.01703.

[43] F. Pedregosa *et al.*, "Scikit-learn: Machine Learning in Python," *J Mach Learn Res*, vol. 12, no. null, pp. 2825–2830, Nov. 2011.

[44] J. Schreiber, "Pomegranate: fast and flexible probabilistic modeling in python," *J Mach Learn Res*, vol. 18, no. 1, pp. 5992–5997, Jan. 2017.

[45] "End-Use Load Profiles for the U.S. Building Stock." Accessed: Mar. 14, 2024. [Online]. Available: https://www.nrel.gov/buildings/end-use-load-profiles.html

[46] C. Wang, D. Yan, and Y. Jiang, "A novel approach for building occupancy simulation," *Build. Simul.*, vol. 4, no. 2, pp. 149–167, Jun. 2011, doi: 10.1007/s12273-011-0044-5.

[47] N. Luo *et al.*, "A three-year dataset supporting research on building energy management and occupancy analytics," *Sci. Data*, vol. 9, no. 1, Art. no. 1, Apr. 2022, doi: 10.1038/s41597-022-01257-x.

[48] "Hourly Building Monitoring: Home." Accessed: Nov. 29, 2023. [Online]. Available: https://benchmark8760.org/

[49] "Standard 90.1." Accessed: Jan. 29, 2025. [Online]. Available: https://www.ashrae.org/technical-resources/bookstore/standard-90-1

[50] D. Koller and N. Friedman, *Probabilistic Graphical Models: Principles and Techniques - Adaptive Computation and Machine Learning*. The MIT Press, 2009.

[51] D. P. Kingma and M. Welling, "Auto-Encoding Variational Bayes," May 01, 2014, *arXiv*: arXiv:1312.6114. doi: 10.48550/arXiv.1312.6114.

# APPENDICES

## Appendix A: Occupancy Inference as a Hidden State Inference Task – Detailed Formulation

### 1. Problem formulation using ELBO

Recall the definition of the key variables and parameters in this hidden state inference task:
- $P$: Observed outputs, specifically the load metering data.
- $Z$: Hidden or latent states, representing the underlying occupancy levels.
- $X$: Exogenous variables influencing the observed output $Y$ and (or) the hiddeb state $Z$, including environmental factors such as outdoor air temperature and calendar variables such as hour of the day.
- $\theta$: Model parameters, reflecting how $Y$ is influnced by $Z$ and $X$. These factors may correspond to physical properties of energy systems. For instance, they may stand for installed capacity of lighting and plugs, or heat loss factors relevant to HVAC systems).

The goal can be described as Maximum Likelihood Estimation (MLE) under hidden states, i.e., to maximize the log-likelihood $\log p(P \mid X; \theta)$ with respect to $\theta$, where the hidden states $Z$ are unknown and must be inferred together with the estimation of $\theta$. Now, directly computing the marginal likelihood $p(P \mid X; \theta)$ becomes intractable due to the unknown $Z$. In fact, inferring hidden state is equivalent to obtain the complete posterior distribution of $Z$, $p(Z \mid P, X; \theta)$, for each observed $Y$ and associated $X$. We would also be able to address this MLE under hidden state task as long as $p(Z \mid P, X; \theta)$ is available. Here we introduce the well-established approach that introduce another surrogate distribution $q(Z \mid P, X)$, which is a distribution of $Z$ that can vary with $Y$ and $X$ that aims to approximate the desired posterior $p(Z \mid P, X; \theta)$, and reformulate the problem with Evidence Lower Bound (ELBO) [1]. Specifically, we transform the log likelihood of $p(P \mid X; \theta)$ using $q(Z \mid P, X)$:

$$\log p(P|X;\theta) = \log \frac{p(P, Z|X;\theta)}{p(Z|P, X;\theta)} = \log \frac{p(P, Z|X;\theta)}{q(Z|P, X)} - \log \frac{p(Z|P, X;\theta)}{q(Z|P, X)}$$

Then, we multiply $q(Z \mid Y, X)$ on both sides and integrate over $Z$:

$$\int q(Z|P, X) \log p(P|X;\theta) \, dZ = \int q(Z|Y, X) \log \frac{p(P, Z|X;\theta)}{q(Z|P, X)} \, dz - \int q(Z|Y, X) \log \frac{p(Z|P, X;\theta)}{q(Z|Y, X)} \, dz$$

After integral, the left-hand side changes back to the original log-likelihood term. On the right-hand side, we get an ELBO and a Kullback-Leibler (KL)-divergence term:

$$\log p(P|X;\theta) = \underbrace{\int q(Z|P, X) \log \frac{p(P, Z|X;\theta)}{q(Z|P, X)} \, dz}_{\text{ELBO}} + KL\big(q(Z|P, X) | p(Z|P, X;\theta)\big)$$

(A.1)

Here, the KL term measures the "distance" between the surrogate $q(Z|P, X)$ and the actual hidden state posterior distribution $p(Z \mid P, X; \theta)$, which is always non-negative. Therefore, the ELBO is always equal or less than the log-likelihood objective, as its name indicates. This ELBO formulation unnifies the explanations of almost all the hidden state inference methods. Here, we just brief the most influential two algorithms tha inspire our model design: Expectation-Maximization (EM) and Variational Auto-encoding (VAE).

### 2. EM algorithm

As its name indicates, the EM algorithm alternates between two steps: inferring the hidden states $Z$ (Expectation step) and estimating $\theta$ (Maximization step). In the EM algorithm, while the exact parameters of $p(P \mid X; \theta)$ and $p(P, Z|X; \theta)$, are unknown, it is crucial that their explicit functional forms are specified in advance. Specifically:

- E-step (Expectation): in this step, using the parameter estimate $\theta^*$ from last iteration, rom the previous iteration, we set the surrogate distribution $q(Z|P, X)$ exactly as the posterior distribution $p(Z|P, X; \theta)$. In this case, the KL term shrinks to zero. As a result, As a result, the ELBO becomes identical to the log-likelihood, according to Equation (A.1). This equivalence means that maximizing the ELBO with respect to $\theta$ is the same as performing MLE of $\theta$, given the current distribution of $Z$. By examining the ELBO more closely, we observe that only part of it depends on $\theta$ and needs update. This part is the expectation of the joint log-likelihood under the surrogate distribution $q(Z|P, X)$, commonly denoted as $Q(\theta)$, as shown in Equation (A.2). This is also the "Expectation" term referred in its name:

$$Q(\theta) = \int q(Z|P, X) \log p(P, Z|X; \theta) = E_{q(Z|P, X)}[\log p(P, Z|X; \theta)]$$
(A.2)

- M-step (Maximization): then, in this step, the goal is to identify the parameter $\theta^*$ that maximizes the expectation term $Q(\theta)$. After determining the updated parameter $\theta^*$, the algorithm proceeds by repeating the E-step. By alternating between these two steps, the updates to $Q(\theta)$ are guaranteed to increase the log-likelihood at each iteration. As a result, the EM algorithm is theoretically guaranteed to converge.

The EM algorithm is widely used in many applications, particularly in clustering methods. For example, models such as Gaussian Mixture Models (GMMs) and Hidden Markov Models (HMMs), which aim to recover underlying clusters or components as hidden states, are traditionally solved using the EM algorithm. Also, as mentioned in Section 3.4.1 in the main chapters, we use the same less form as $Q(\theta)$ when updating the model parameters in the load disaggregator.

*3. Variational inference and VAE*

However, in many cases, it is often impossible to explicitly define $p(P \mid X; \theta)$ and $p(P, Z|X; \theta)$ due to their complex or high-dimensional nature. In such cases, variational inference provides a promising alternative. Instead of alternating between steps as in the EM algorithm, we could parameterize the surrogate distribution $q(Z|P, X)$ using a flexible and tractable form and directly optimizes the ELBO. By maximizing the ELBO, we can obtain a credible estimate of $\theta$ Simultaneously, the surrogate distribution $q(Z|Y, X)$ becomes a close approximation of the true posterior $p(Z|P, X; \theta)$, as minimizing the KL divergence term is inherently tied to maximizing the ELBO (Equation (A.1)).

In line with this concept, VAE [2] leverages neural networks—the most flexible and powerful tools for function approximation—to represent the surrogate distribution $q(Z|P, X)$. VAE and its numerous variants are now powerful tools for hidden state inference tasks involving complex and high-dimensional data, such as compressing images or videos and capturing semantic information from them. Additionally, the development of VAE models also explored many strategies to incorporate prior in the hidden state space in the deep learning framework, For example, in Variational Quantized-VAE (VQ-VAE) [3], the hidden space is represented by a finite number of candidate samples as possible posterior hidden states, which offers a practical alternative when the prior knowledge cannot be explicitly defined.

*4. Special challenges of occupancy inference*

Although EM and VAE are well-established for general hidden state inference tasks, inferring occupancy from energy metering data presents unique challenges:
1. The distribution of $Z$ (occupancy) is heterogeneous. Unlike the universal hidden state distributions often assumed in traditional inference tasks, office building occupancy patterns vary significantly throughout the day, reflecting the diurnal rhythm of social activities.
2. On top of this, the impacts of exogenous variables are complex. For example, besides that occupancy influences HVAC loads, outdoor weather conditions also exert significant non-linear effects.

Traditional clustering methods often rely on strong assumptions of homogeneous hidden state distributions and linear impacts of exogenous variables, limiting their effectiveness in our specific case. In contrast, neural networks are inherently powerful tools for capturing both non-linearity and heterogeneity, enabling them to learn effective representations of observed energy output. However, without explicit interpretability or control, it is difficult to link these representations to the underlying occupancy—a critical requirement for downstream occupant-centric operations. Moreover, the black-box nature of neural networks remains a significant challenge, which raises concerns about their robustness and generality, especially when applied to typically limited building energy metering data.

To address these challenges, our model builds on the general hidden state inference framework but includes two key components tailored to our specific problem. First, an occupancy profile generator incorporates prior knowledge to reflect and control the realistic heterogeneity of office occupancy patterns. Second, a controllable and interpretable load disaggregator, implemented as small neural network layers, captures the complex impacts of both occupants and environmental factors on different energy systems while remaining consistent with established building energy behavior.

*Appendix B: Generating Candidate Stochastic Posterior Distribution Series from reference schedules*

To represent the heterogeneity of occupancy patterns over a day, we generate time-indexed distributions, each representing a candidate posterior occupancy ratio $p(Z_t|Y_t, X_t)$ at time step $t$. As discussed in Section 3.2.1 in the main chapters, the generated candidate distributions must align with prior knowledge of typical office building occupancy patterns. These patterns often vary across different hours of the day and between day types, such as weekdays and weekends. A practical way to incorporate this prior knowledge is by introducing controlled disturbances to deterministic reference occupancy schedules. The reference schedules should at least cover one typical working day and one non-working day and may contain more typical scenarios. These schedules can be from various sources, such as local design guides or insights from building managers. The sources of reference schedules will be discussed in Section 5.

To introduce controlled disturbances to these reference schedules, we first compute a metric $a_{t,i}$, called the level-distance score. This score quantifies the deviation of the fractional occupancy ratio in the reference schedule, $r_t^{\text{ref}}$, from the assigned centroid in the categorical distribution, $c_j$, at each time step $t$:

$$a_{t,j} = |r_t^{ref} - c_j|$$

(B.1)

To generate samples with varying degrees of dispersion in the distributions, we introduce a random variable called "temperature", denoted as $\tau$. to generate various samples with different degrees of dispersion in the distributions. The level-distance scores $a_{t,i}$ are divided by the temperature $\tau$ to obtain the logits $s_{t,i}$:

$$s_{t,j} = -\frac{a_{t,j}}{\tau}$$

(B.2)

Here, $\tau > 0$ governs the sharpness of the resulting distribution. Larger $\tau$ produces a highly dispersed distribution where probability masses are more uniform across all levels, while lower $\tau$ leads to a more peaked distribution where one level dominates. Additionally, prior knowledge of office occupancy patterns suggests that variability in occupancy levels changes throughout the day. To reflect this variability, we model $\tau$ as a uniform random variable over the range $[0, T_t]$, where the upper bound $T_t$ depends on the time of day. Specifically, during inactive periods, such as nighttime and weekends, occupancy patterns are highly predictable and typically close to zero. For these periods, we assign a low value to $\tau$ to reflect the low uncertainty in occupancy levels. Conversely, during daytime hours on working days, occupancy patterns are more variable, with peak and valley counts fluctuating day to day. For these periods, we assign a much larger $\tau$, allowing for more dispersed and uncertain distributions.

By sampling multiple values of $\tau$ from its period-dependent distribution, we generate corresponding logits for each sample.

Next, the logits $s_{t,j}$ are normalized using the softmax function to construct a valid categorical distribution, as shown in Equation (B.3). Finally, these probabilities are used to generate numerous time-indexed stochastic samples of the ordered categorical distribution. Since reference schedules and the distribution of $\tau$ may differ across day types, the samples are generated as daily series, resulting in probabilistic daily occupancy profiles, as shown in Figure 3. The aforementioned GM proxy is also constructed accordingly, with each level's probability determining the weight of its associated Gaussian component.

$$P_{t,j} = \frac{exp(s_{t,j})}{\sum_{j=1}^{N} exp(s_{t,j})} \tag{B3}$$

**REFERENCES**
[1] D. Koller and N. Friedman, *Probabilistic Graphical Models: Principles and Techniques - Adaptive Computation and Machine Learning*. The MIT Press, 2009.
[2] D. P. Kingma and M. Welling, "Auto-Encoding Variational Bayes," May 01, 2014, *arXiv*: arXiv:1312.6114. doi: 10.48550/arXiv.1312.6114.
[3] A. Razavi, A. van den Oord, and O. Vinyals, "Generating Diverse High-Fidelity Images with VQ-VAE-2," Jun. 02, 2019, *arXiv*: arXiv:1906.00446. doi: 10.48550/arXiv.1906.00446.